\documentclass{article}
\usepackage{arxiv}

\usepackage{hyperref}
\usepackage{bm}
\usepackage{cite}
\usepackage{graphics,fancyhdr,graphicx,subfigure}
\usepackage{subfigure}
\usepackage{amsthm,amsmath,amsfonts,latexsym,amssymb}
\usepackage{lineno}
\usepackage{diagbox}
\usepackage{algorithm}
\PassOptionsToPackage{noend}{algpseudocode}
\usepackage{algpseudocode}
\usepackage{multirow,booktabs}
\usepackage{listings}
\usepackage[table]{xcolor}
\usepackage{lscape}
\usepackage{comment}
\usepackage{tabularx}
\usepackage{graphicx}
\usepackage{tabularx}
\usepackage{nomencl}
\usepackage{array}
\usepackage{caption}
\usepackage{breqn}
\usepackage{lineno}
\usepackage{mathtools}
\usepackage{subfigure}
\usepackage{caption}
\usepackage{lmodern}
\usepackage{calligra}
\usepackage{xspace}
\usepackage[utf8]{inputenc}
\usepackage{enumerate}
\usepackage[english]{babel}

\def\equationautorefname~#1\null{%
  Eq.~(#1)\null
  }
\def\subfigureautorefname~#1\null{%
  Fig.~#1\null
}

\definecolor{listinggray}{gray}{0.9}
\definecolor{lbcolor}{rgb}{0.9,0.9,0.9}
\definecolor{Darkgreen}{RGB}{0,100,0}

\title{Stochastic projection based approach for gradient free physics informed learning \thanks{https://www.csccm.in/}}

\author{ \hspace{1mm}Navaneeth~N.\\
	Department of Applied Mechanics\\
	Indian Institute of Technology (IIT) Delhi\\
	Hauz Khas - 110 016, New Delhi, India \\
	\texttt{navaneeth.n@am.iitd.ac.in} \\
	\And
	\hspace{1mm}Souvik~Chakraborty \\
	Department of Applied Mechanics\\
	Indian Institute of Technology (IIT) Delhi\\
	Hauz Khas - 110 016, New Delhi, India \\
	\texttt{souvik@am.iitd.ac.in} \\
}

\renewcommand{\shorttitle}{Koopman operator for time-dependent reliability analysis}

\hypersetup{
pdftitle={Stochastic projection based approach for gradient free physics informed learning},
pdfsubject={q-bio.NC, q-bio.QM},
pdfauthor={Navaneeth N., Souvik Chakraborty},
pdfkeywords={Stochastic Projection, Physics informed neural network, Partial differential equations, Automatic differentiation, loss function},
}

\begin{document}
\maketitle

\begin{abstract}
We propose a stochastic projection-based gradient free physics-informed neural network. The proposed approach, referred to as the stochastic projection based physics informed neural network (SP-PINN), blends upscaled stochastic projection theory with the recently proposed physics-informed neural network. This results in a framework that is robust and can solve problems involving complex solution domain and discontinuities. SP-PINN is a gradient-free approach which addresses the computational bottleneck associated with automatic differentiation in conventional PINN. Efficacy of the proposed approach is illustrated by a number of examples involving regular domain, complex domain, complex response and phase field based fracture mechanics problems. Case studies by varying network architecture (activation function) and number of collocation points have also been presented.  
\end{abstract}


\section{Introduction} \label{sec:intro}
Physical and real world phenomenons are often represented by Partial Differential Equations (PDEs). Unfortunately, solving PDEs employing analytical methods is possible only for a limited class of problems and hence, development of numerical methods/schemes for solving PDEs have gained wide attention over the past few decades. Popular numerical methods for solving PDEs include finite element method (FEM) \cite{aruoba2003finite}, finite difference method (FDM) \cite{khabaza2014numerical}, finite volume methods (FVM) \cite{leveque2002finite} and spectral methods \cite{kopriva2009implementing}. However, the existing numerical methods are often computationally prohibitive, especially for solving problems involving design and optimization, and parameter estimation, uncertainty quantification and propagation, and reliability analysis. Therefore, development of efficient methods for solving PDEs is still a relevant area of research in computational science and engineering.

\par
Advancements in machine learning (ML) \cite{lin2021uncertainty,surrogate2022} and Deep Learning (DL) \cite{de2022long}-based approaches have started revolutionizing various engineering fields, and computational mechanics is no exception. A plethora of machine learning based algorithms including operator learning \cite{lu2021learning,garg2022assessment,garg2022variational,tripura2022wavelet} and physics-informed neural network (PINN) \cite{baker2019workshop, raissi2019physics, goswami2020transfer,chakraborty2021transfer,karumuri2020simulator} have been developed over the past few years. While operator learning  attempts to learn the solution of parametric PDE from data, PINN aims at training a neural network model by directly satisfying the governing physics. In this paper, our aim is to devise a novel physics-informing learning framework, and hence, the discussion hereafter is confined to PINN only. Readers interested in exploring operator learning can refer \cite{tripura2022wavelet,thakur2022multi,lu2021learning}.


\par
Although the term PINN was coined in the seminal work by Raissi \textit{et al.} \cite{raissi2019physics}, the concept was originally proposed in early 1990s \cite{dissanayake1994neural}. Over the past few years, PINN has emerged as a viable alternative for solving PDEs. Training deep neural network is computationally intensive and can easily lead to memory constraints, and physics-informed neural network is no exception. In order to surmount the hurdle, Amabathiran et al. proposed a Sparse PINN (SPINN) \cite{ramabathiran2021spinn}. The fundamental idea here is to employ  a sparse Deep Neural Network (DNN), which reinterprets the traditional meshless representation of PDE. Similarly, \cite{sirignano2018dgm} presented Galerkin projection-based meshless deep neural network framework \cite{sirignano2018dgm} capable of solving high-dimensional PDEs. Apart from the memory limitation, one of the other major drawbacks of the PINN is inefficient training and low accuracy. Upon identification of the shortcoming, Chiu et al.  \cite{chiu2022can} proposed Coupled-Automatic-Numerical PINN (CAN-PINN) which couples the automatic differentiation with the numerical differentiation. 
\cite{kharazmi2021hp} proposed a hp adaptive variational PINN (hp-VPINN) and illustrated its efficacy in solving PDEs. Other significant work in this area includes  \cite{jagtap2020conservative,jagtap2020extended,wandel2020learning,wandel2020teaching,lu2021physics,gao2021phygeonet,sharma2022accelerated}.

\par
Despite the fact that PINN models have proven to be a potential solution for a broad variety of physical phenomena, the model accuracy and training efficiency often remain challenging. 
For instance, a large number of collocation point might be needed for solving PDEs with nonsmooth solutions \cite{mcclenny2020self,nabian2021efficient}. This inherently results in increased computational cost. On the other hand, PINN is over-parameterized and hence, training with insufficient collocation points leads to unreliable solutions. Researchers \cite{xiang2022hybrid,chiu2022can} have attempted to address this by combining numerical differentiation with automatic differentiation; however, selection of suitable numerical differentiation schemes remains an open-problem in this regards.  


Another major hurdle is associated with the choice of network architecture. PINN generally utilizes a fully connected network that constrains the physics through residual loss calculated at the collocation points. Since the training loss is computed through Automatic Differentiation (AD) \cite{baydin2018automatic}, the network must be differentiable in accordance with the requirements of the governing PDE. This can potentially restrict our choice and
prohibit using some of the most popular activation functions including ReLU and LeakyReLU \cite{lee2020correctness}.

In this paper, we present a gradient-free PINN, referred to as the Stochastic  Projection based PINN or SP-PINN, for solving PDEs. The proposed approach blends stochastic projection theory with PINN and eliminates the need for gradient computation. Specifically, we employ filtering within the scope of stochastic projection theory to compute the gradients. The proposed framework is mesh-free and falls under the broad umbrella of particle based methods. The proposed SP-PINN is robust and is ideally placed for solving problems involving non-smooth solution, non-trivial domains and discontinuities. Moreover, the proposed approach strives to deliver accurate predictions efficiently, even with limited collocation points.



The remaining sections are organized as follows. In Section  \ref{sec:pbs}, the technical context of the problem is described. Detailed description of the proposed SP-PINN is discussed in
Section \ref{sec:pa}. In Section \ref{sec:ne}, numerical examples illustrating the proposed methodology are provided. Lastly, Section \ref{sec:conclusions} provides the concluding remarks.
\section{Technical context of the problem}\label{sec:pbs}
The general aim of this work is to develop SP-PINN and illustrate its utility in solving PDEs. For that, we consider a residual form of a general parameterised  non-linear partial differential equation;
\begin{equation}\label{genPDE}
\begin{aligned}
&\mathcal{N} \left(\bm{x}, t, {u}, \partial_{\mathbf{t}} {u},\partial^2_{\mathbf{t}} {u}\ldots , \partial_{x} {u}, \partial^n_{t} {u}, \ldots,\partial^n_{x} {u}, \boldsymbol{\alpha}\right)=0, \quad \bm{x} \in \Omega, t \in[0, T] 
\end{aligned}
\end{equation}
with initial conditions and boundary conditions respectively of the forms,
\begin{equation}\label{genBC}
\begin{aligned}
&{u}\left(\bm{x}, 0\right)=g_{0}(\bm{x}) \quad \bm{x} \in \Omega, \\
&\mathcal{B}[{u}(\bm{x}, t)]=g_{\Gamma}(t), \quad \bm{x} \in \partial \Omega,\, t \in[0, T].
\end{aligned}
\end{equation}
Here, $t$ represents the time, while $bm{x} \in \mathbb{R}^{d}$ represents the spacial coordinate. $\mathcal{N}$ in Eq. \eqref{genPDE} represents a nonlinear operator and is parameterized by $\bm \alpha$.  
In theory, it is possible to solve Eq. \eqref{genPDE} by using PINN \cite{raissi2019physics}; however, in practice this is not straightforward specifically for problems involving complex domain, discontinuities and non-smooth solution. Additionally, AD used within vanilla PINN is a major computational bottleneck. The objective here is to develop a variant of PINN that addresses the aforementioned challenges. 

\section{Proposed methodology}\label{sec:pa}
In this section, we present the proposed framework. However, before delving into the details of the proposed approach, we briefly discuss the stochastic projection method. 

\subsection{Derivative through stochastic projection method}
Stochastic projection method is inspired by the modelling of continuum-atomistic kinematics. The notion of a directional derivative is achieved by employing a probabilistic projection method exploiting microstructural information in a multi-scale formulation of a continuum body. Thus, the stochastic projection principle can be regarded as an up-scaling of atomistic information that yields non-trivial directional information to evolve the state variables. 
As the theory implies, the stochastic projection principle extends the definition of directional derivative and can be regarded as a discrete Cauchy-Born map forming a classical deformation gradient in the infinitesimal limit \cite{sunyk2003higher}. Similarly, the work \cite{nowruzpour2019derivative} discusses the preservation of micro-scale information when a continuum body undergoes deformation as a multi-scale process namely  macro-micro scales. 

To discuss the stochastic projection in perspective of physics-informed learning, we consider a domain with finite number of collocation points. The field over the domain varies according to the boundary condition. Now, for every collocation point, there exists a neighborhood specified by the distance. Variation of the field variable is distinctively measured till certain characteristic distance, and can be considered as macroscopic-scale. On the other hand, variation of the field variable less than the characteristic length can be considered to be microscopic scale. Similarly, the field variables can be differentiated into those which evolve slowly with time and those which evolve on faster time scales based on the characteristic time interval. When an observer measures the variation in the the macroscopic level, the fluctuations in the microscopic levels may be treated to be a stochastic process, as otherwise it remains unaccounted. 
Now, suppose $\bm u(\bm x)$ to be a macroscopic field variable, prior
to conditioning based on any microscopically inspired information, then a zero mean noise term is added to the $\bm u(\bm x)$ to obtain $\bm u(\bm z)$, which can be expressed as:
\begin{equation}
    \bm u(\bm z) = \bm u(\bm x)+\mathbf \Delta \eta
\end{equation}
where $\bm z \neq \bm x$, and $\bm u(\bm z)$ is the field measurement at $\bm z$. Here, $\mathbf \Delta \eta$ accommodates the unaccounted fluctuations in the microscopic level (noise). Thus, the problem statement deduce to an attempt of characterizing  macroscopic field variable $\bm u(\bm x)$ by means of neighborhood information. After conditioning based on the information from the micro scale, sampled at time $t$, the noisy observation at the macroscopic level can be written in the following form;
\begin{equation}\label{sampled observation}
    d{\bm {Z}_t} =\bm{h}(\bm{x}_t,\bm{z}_t)d{t}+\sigma{d{\bm{W}_t}}.
\end{equation}
In the above expression, $\bm {h}(\cdot,\cdot)$ returns the difference in the field variable at points located at a distance which can be measured on macroscopic scale. The term $\sigma{d{\bm{W}_t}}$ represents the noise which depends on its microscopic counter part and microscopically sampled function $\bm{Z}_t$. Further a conditional expectation of $\bm u$ is utilised to obtain the microscopically informed spacial variation in the field variable, which is given by: 
\begin{equation}
    \Pi_t(\bm u) = E_{\mathcal P}[\bm u(\bm x)|\mathcal F_t].
\end{equation}
Here $\mathcal F_t$ represents the sequence formed by adding up $\bm Z_t$ at each time till t, and $\bm {W}_t$ is the Brownian motion independent of $\boldsymbol{\eta}_t$. Also, note that the time $t$ is restricted to a given characteristic microscopic time interval $\delta \hat{t}$. Hence, the problem can be further considered to be an analogy of stochastic filtering, having a process measurable in the probability space $(\Omega, \mathcal{F}, \mathcal P)$. Now to describe the conditional distribution under a different probability measure $\mathcal Q$, where $\bm Z(t)$ behaves as drift removed Brownian motion, we utilize Kallianpur-Stribel formula \cite{kallianpur1968estimation}, where $\mathcal Q$ is continuous with respect to $\mathcal P$. According to the formulation of change of probability measure \cite{girsanov1960transforming}, conditional distribution of a continuous and twice differentiable function, $\phi$ can be expressed as 
\begin{equation}
\pi_{t}(\phi)=\frac{\sigma(\phi)}{\sigma(1)}:=\frac{E_{\mathcal Q}\left[\phi_{t} \Lambda_{t} \mid \mathcal{F}_{t}\right]}{E_{\mathcal Q}\left[1 \Lambda_{t} \mid F_{t}\right]},
\end{equation}
with $\Lambda_t$ as the Radon-Nikodym \cite{kallianpur2013stochastic} derivative used in the change of measure. The expression of $\Lambda_t$ is given by
\begin{equation}
\Lambda_{t}=\frac{d \mathcal P}{d \mathcal Q}=\exp \left(\sum_{i}\left(\int_{0}^{t} \bm {h}^{t}\left(\bm{x}_{s}\right) d \bm{z}_{s}^{{i}}-\int_{0}^{t} \bm{h}^{i}\left(\bm{x}_{s}\right)^{2} d s\right)\right).
\end{equation}
The stochastic differential equation (SDE) pertaining the evolution of $\Lambda_{t}$ can be expressed as:
\begin{equation}
d \Lambda_{t}=\Lambda_{t} \bm{h}_{t}^{T} d Z_{t}.
\end{equation}
Further, time evolution equation for the normalized conditional law $\pi_t{\phi}$ takes the following form:
\begin{equation}\label{eq9}
d \pi_{t}(\phi)=\left(\pi_{t}\left(\phi \bm{h}^{T}\right)-\pi_{t}(\bm{h})^{T} \pi_{t}(\phi)\right)\left(\sigma \sigma^{T}\right)^{-1}\left(d \bm{z}_{t}-\pi_{t}(\bm{h}) d t\right).
\end{equation}
Replacing $\phi=\bm z - \bm x$ in \autoref{eq9} leads to the equation:
\begin{equation}\label{eq10}
d \pi_{t}(\bm{z}-\bm{x})=\left(\pi_{t}\left((\bm{z}-\bm{x}) \bm{h}^{T}\right)-\pi_{t}(\bm{h})^{T} \pi_{t}(\bm{z}-\bm{x})\right) \cdot\left(\sigma \sigma^{T}\right)^{-1}\left(d \bm{z}_{t}-\pi_{t}(\bm{h}) d t\right)
\end{equation}
We assume that initially, the field variable over the domain remains the same for all collocation points. As previously stated, the macroscopic field variable cannot be resolved into a vector having length less than the characteristic length, $|\Delta|$; therefore, \autoref{eq10} takes the following form for  $\bm z_{t_0}-\bm x_{t_0}=\Delta$ with $t_0$ being the initial time:
\begin{equation}\label{eq11}
    (\bm{z}_t-\bm{x}_t)= (\bm{z}_{t_0}-\bm{x}_{t_0})+  \int_{\hat{t_0}}^{\hat{t}} \left(\pi_{s}\left((\bm{z}-\bm{x}) \bm{h}^{T}\right)-\pi_{s}(\bm{h})^{T} \pi_{t}(\bm{z}-\bm{x})\right) \cdot\left(\sigma \sigma^{T}\right)^{-1}\left(d \bm{z}_{t}\right)
\end{equation}
Here, $\bm h(\bm x, \bm z)$ is chosen such that it is smooth enough and satisfies
\begin{equation}
    \bm h(x,z) =
    \begin{cases}
      0, & \text{if}\ |z-x|\leq 0 \\
      Nonzero, & \text{otherwise}
    \end{cases}
\end{equation}
It is worthwhile to note that in \autoref{eq11}, $\bm Z_t$ represents the specific observation: $\bm Z_t = \bm Z_{t^'} + \int _{\delta \hat{t}}{\Delta}{d\hat{s}}$ with $t^{'}$ as the previously sampled macroscopic time. From the \autoref{sampled observation}. variance of the macroscopic observation is obtained empirically as ${\boldsymbol{\sigma}} {\boldsymbol{\sigma}}^T = \pi_t(\bm h \bm h^{T})\delta \hat{t}$.

Macroscopic level temporal fluctuations in the integrand are not solvable since the integration is performed within the least microscopically resolvable time period $t$ and hence, $\bm z_t -\bm x_t$ is drift less. Hence the expression of $\bm z_t - \bm x_t$ can be approximated as:
\begin{equation}
    (\bm{z}_t-\bm{x}_t)\sim \Delta + \mathbf{G}\Delta
\end{equation}
where the expression of $\mathbf{G}$ is given by 
\begin{equation}
    \mathbf G = \left(\pi_{t}\left((\bm{z}-\bm{x}) \bm{h}^{T}\right)- \pi_{t}(\bm{z}-\bm{x})\right)\pi_{t}(\bm h)^{T})((Var(\bm h))^{-1}.
\end{equation}
It is imperative to note here that $\mathbf G$ utilizes the non-infinitesimal neighborhood information to compute the gradient. The key feature of the gradient-free approach includes
\begin{itemize}
    \item The gradient is achievable even when there are discontinuities in the field.
    \item Though the measure of $\mathbf{G}$ is problematically founded, the integral expression of $\mathbf{G}$ at a given point ($\bar{\bm X}$) can be quantitatively assessed by employing the Monte-Carlo approach from the $N_t$ number of neighborhood points
    \begin{equation}\label{spgf}
    \mathbf{G}(\bm X=\bar{\bm X}) = \frac{\partial u}{\partial \bm X}= \frac{\frac{1}{N_t}{\sum}_{i=1}^{N_{b}}{(u-\bar{u})(\bm X_i-\bar {\bm X})^{T}}}{\frac{1}{N_t}{\sum}_{i=1}^{N_{b}}{(\bm X_i-\bar{\bm X})(\bm X_i-\bar {\bm X})^{T}}}
    \end{equation}
\end{itemize}
\subsection{Gradient Free Physics informed Neural network}\label{sec:gradeint Free}
While in the  previous section we have discussed the stochastic projection-based gradient in detail, in this section, we explain the integration of the gradient-free approach with the neural network to obtain  a Stochastic Projection-based Physics Informed Neural Network (SP-PINN) . Before going into the details of the proposed approach, we briefly review the commonly employed PINN \cite{raissi2019physics}. To that end, we consider a governing equation of the form; $\mathcal{N} \left(\bm{x}, t, {u}, \partial_{\mathbf{t}} {u},\partial^2_{\mathbf{t}} {u}\ldots , \partial_{x} {u}, \partial^n_{t} {u}, \ldots,\partial^n_{x} {u}, \boldsymbol{\alpha}\right)=0 = 0$ with $u$ as field variable as given in the \autoref{genPDE} . Now to represent the PINN, a fully connected feed-forward network is utilised, i.e $\mathcal{U}(\bm x,t,\bm{w})$. The residual form can be rewritten in terms of the differential operators, $\mathcal{N_{\bm x}}$ and  $\mathcal{N_{\bm x}}$, such that $\mathcal{N}_{t}[u(\bm x, t)],\, \mathcal{N}_{\bm x}[u(\bm x, t)]$ are the terms composed of derivatives with respect to the temporal and spatial variables respectively:
\begin{equation}
\mathcal{N}_{t}[\mathcal{U}(\bm x, t)]+\mathcal{N}_{\bm x}[\mathcal{U}(\bm x, t)]=0.
\end{equation}
Here, the initial condition is given by $\mathcal{U}(\bm x,0) = u_{0}(\bm x), \, x \in \Omega$ and boundary condition is given by $\mathcal{B}[\mathcal{U}(\bm x, t)] = g_{0}(\bm x, t), \, x \in \partial \Omega, \, t \in(0, T]$. The boundary operator, $\mathcal{B}$, represents either Dirichlet boundary, Neumann boundary condition $g(x,t)$ or mixed form. Additionally, in order to train the network, an appropriate loss function is required. Typically, the total loss function to train the PINN is defined as  the sum of the three loss components namely PDE loss ($\mathcal{L}_{PDE}$), boundary loss $\mathcal{L}_{BC}$ and initial condition loss $\mathcal{L}_{IC}$. The components of the loss functions are mathematically expressed as 
\begin{equation}
\begin{aligned}
&\mathcal{L}_{PDE}=\left\|\mathcal{N}_{t}[\mathcal{U}](\cdot ; \bm w)+\mathcal{N}_{x}[\mathcal{U}(\cdot ; \bm w)]\right\|_{\Omega \times(0, T]}^{2}, \\
&\mathcal{L}_{B C}=\|\mathcal{B}[\mathcal{U}(\cdot ; \bm w)]-g(\cdot)\|_{\partial \Omega \times(0, T]}^{2}\\
&\mathcal{L}_{I C}=\left\|\mathcal{U}(\cdot, 0 ; \bm w)-u_{0}\right\|_{\Omega}^{2}
\end{aligned}
\end{equation}
Thus the total loss function is given by:
\begin{equation}
   \mathcal{L}_{Total} = \lambda_{PDE} \mathcal{L}_{PDE} + \lambda_{BC} \mathcal{L}_{BC} + \lambda_{IC} \mathcal{L}_{IC}
\end{equation}
The coefficients $\lambda$s are chosen such that a better convergence of the total loss is achieved. For computing the PDE loss ($\mathcal{L}_{PDE}$), a finite set of collocation points ($n$) are generated in the domain such that $D_{PDE} = \{(\bm x_{i}, t_{i})\}^{n}_{i=1} \in \Omega \times (0,T]$ , where the sampling of points can be either uniformly or non-uniformly distributed. Once the collocation points are generated, derivatives of the network, $ \partial_{\mathbf{\bm x}} \mathcal{U}(\bm x,t; \bm{w}), \partial^2_{\mathbf{t}} \mathcal{U}(\bm x,t; \bm{w}) \ldots \partial_{\bm x} \mathcal{U}(\bm x,t; \bm{w}), \partial^2_{t} \mathcal{U}(\bm x,t; \bm{w}), \ldots,\partial^2_{\bm x}\mathcal{U}(\bm x,t; \bm{w})$ are evaluated at each of these points and employing the residual form $\mathcal{L}_{PDE}$ is computed. The loss components corresponding to boundary loss and initial condition loss are evaluated by  matching the PINN output $\mathcal{U}$ against target $u$ over $m$ labeled samples od initial conditions, $D_{IC} = \{(\bm x_{j}, 0)\}^{m}_{j=1} \in \Omega$, and n samples of boundary conditions, $D_{BC} = \{(\bm{x}_{\partial \Omega}, t_i)\}^{n}_{i=1} \in \partial \Omega \times (0,T]$. Gradient evaluation of the network is essential to obtain the residual component of the loss function. As we stated earlier, the vanilla PINNs \cite{raissi2019physics} utilizes AD to obtain the derivatives field output. On the  contrary, the proposed SP-PINN replaces the AD-based gradients with the stochastic projection-based gradients.  
\par
\begin{figure}
    \centering
    \includegraphics{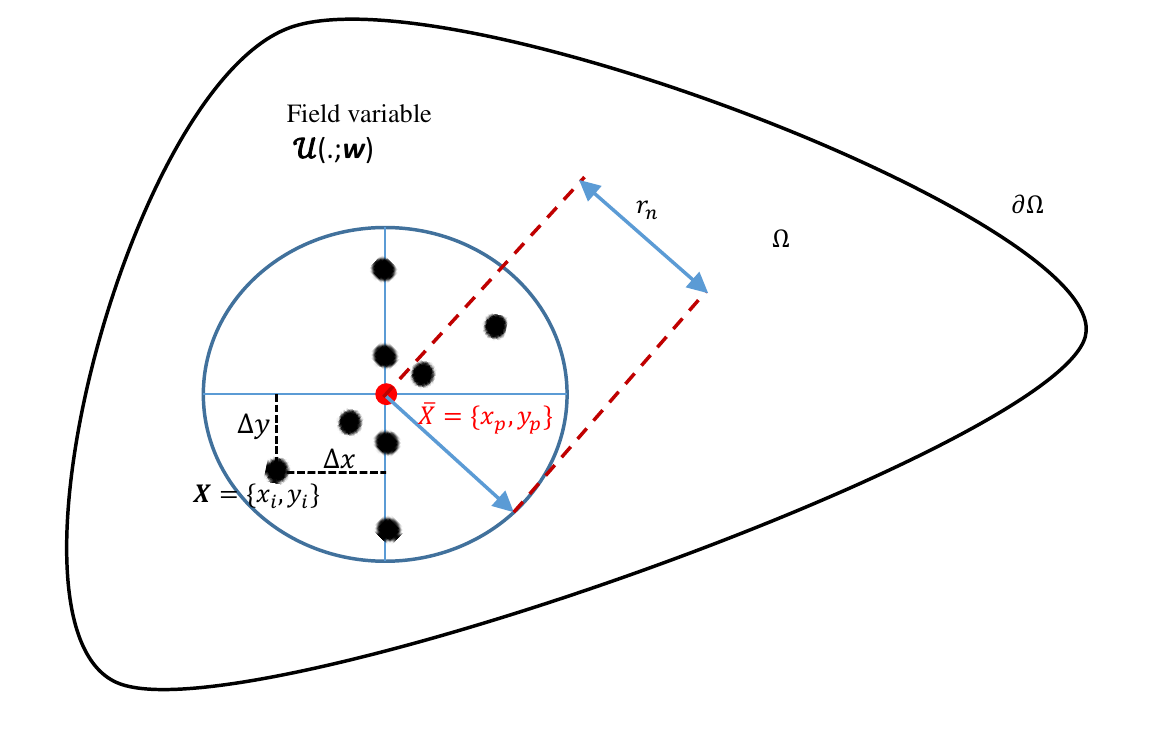}
    \caption{}
    \label{fig:sp_diag}
\end{figure}

To illustrate gradient evaluation of the SP-PINN further, we consider irregular domain, $\Omega$ as shown in \autoref{fig:sp_diag}, where the boundary is defined by $\partial \Omega$. Now to compute the gradients at a given point ${\bm{\bar{x}}} = \{x_p,y_p\}$ in the domain, a neighbourhood is specified in terms of the radius $r_n$. Once the neighborhood is defined, one may choose $N_t$ number of collocation points inside the neighborhood. Subsequently, gradient of network with the input variable at $\bm {\bar{x}}$ is computed by \autoref{spgf_net}, where $\bm{x}_i= \{x_i,y_i\}$ is considered to be a generic neighborhood point. The details of the procedure for implementing SP-PINN is given in Algorithm \ref{alg: Training SP-PINN}.

\begin{equation}\label{spgf_net}
    \mathbf{\hat G}(\bm x={\bm \bar{x}}) = \frac{\partial \mathcal{U}({\bm {\bar x}}, \bm w)}{\partial {\bm {x}}}= \frac{\frac{1}{N_t}{\sum}_{i=1}^{N_{b}}{( \mathcal{U}(\bm {x}_i, \bm w)- \mathcal{U}({\bm \bar{x}}, \bm w))(\bm {x}_i-\bm {\bar {x}})^{T}}}{\frac{1}{N_t}{\sum}_{i=1}^{N_{b}}{(\bm {x}_i-\bm {\bar{x}})(\bm {x}_i-\bm {\bar{x}})^{T}}}
\end{equation}

\begin{algorithm}[t]
\caption{Stochastic Projection based physics Informed Neural Network}\label{alg: Training SP-PINN}
\textbf{Requirements:} Boundary conditions, initial conditions, and PDE describing the physics constrain
\\
{\textbf{Output:} Prediction of the field variable/solution of PDE}
\begin{algorithmic}[1]
\State {\textbf{Initialize:} Network parameters, $\bm{w} = \{w_{i}s, b_{i}s\}$} of the PINN $\mathcal{N}(x,y,t;\theta)$.
\State Generate the collocation  points in the domain $\{x^{i}_{f}, y^{i}_{f} \} \in \Omega$ and $t^{i}_f \in (0,T]$
\State Sample the adequate boundary points in the domain $\{x^{i}_{b}, y^{i}_{b} \} \in \partial \Omega$
\State Define the neighbourhood of the collocation points in terms of distance $\Delta x$ and $\Delta y$
\State Obtain the first order gradients at all the collocation points using \autoref{spgf_net}  and store the gradients
\State From the first order gradients, using the same formulation \autoref{spgf_net} the second order gradients can be achieved. 
\State Define the PDE loss $\mathcal{L}_{PDE}$ in terms of components of the gradients
\State Define the Boundary loss $\mathcal{L}_{BC}$, Sum the all losses to get the total loss $\mathcal{L}_{total}$
\While {{$\mathcal{L} > \epsilon$}}
\State {Train the network:} $\{w_{i}s, b_{i}s\}\leftarrow \{w_{i}s, b_{i}s\}-\delta \nabla_{\bm w, \bm b}{L}(\bm w, \bm b)$
\State {epoch= epoch $+$ 1}
\EndWhile
\State{Return the optimum parameters for the PINN}
\State {Obtain Predictions/solutions}
\end{algorithmic}
\end{algorithm}

\section{Examples}\label{sec:ne}
This section presents numerical illustrations, where we exemplify the proposed method. The examined numerical examples include heat conduction equation, Burgers equation and Poisson's equation. We start by illustrating the performance of the proposed SP-PINN in solving problems involving regular domain, and subsequently proceeds to problems involving non-smooth response, irregular domain, and discontinuity (in form of fracture). 
For each of these problems, we compare the predictions of SP-PINN with actual solution.  Robustness of the proposed approach is evaluated by varying the number of collocation points, number of neighbourhood points, and employing three different activation functions namely ELU, ReLU and tanh. Apart from the aforementioned examples, comparative studies are carried out to evaluate the performance of the SP-PINN and the AD-PINN for the problems with non-smooth solution and problems defined over nontrivial geometries. Lastlty, we demonstrate an example where the solution over the domain has discontinuity in form of a fracture.
In all the above stated examples, implementations of our framework are carried out in python utilising the library \texttt{PyTorch}. All the codes will be made publicly available on acceptance of this paper.
\subsection{Heat conduction Problem}
As the first example, we consider heat conduction equation. The corresponding PDE in the one dimensional space is given as:
\begin{equation}\label{heatconduction}
 u_t = a^2 u_{xx}+ f(x,t) ,\hspace{2em}   x\in [0,l],\hspace{2em}   t\in[0,T],
\end{equation}
with boundary conditions, $u(0,t)=u(l,t)=0$ and initial conditions, $u(x,0)=0$. In \autoref{heatconduction} we assign $a=l=T=1$ and $f(x,t) = {A}sin(\frac{{\pi}{x}}{l})$, where the value of $A$ is set to be 100. Now, in regards to the implementation of the proposed framework, we choose a simple the network architecture with 3 hidden layer withs 40, 80, and 40 neurons. L-BFGS optimizer with learning rate ($l_r$) of $0.1$ is used. For training the neural network, a randomly generated set of boundary and initial  points ($N_b=400$) and equidistant collocation points are utilised. The PDE loss ($L_{PDE}$) at the collocation points is computed through the stochastic projection method described in  Algorithm \ref{alg:Training SP-PINN}. 

\begin{figure}[!h]
    \centering
    \includegraphics[width=.4 \textwidth]{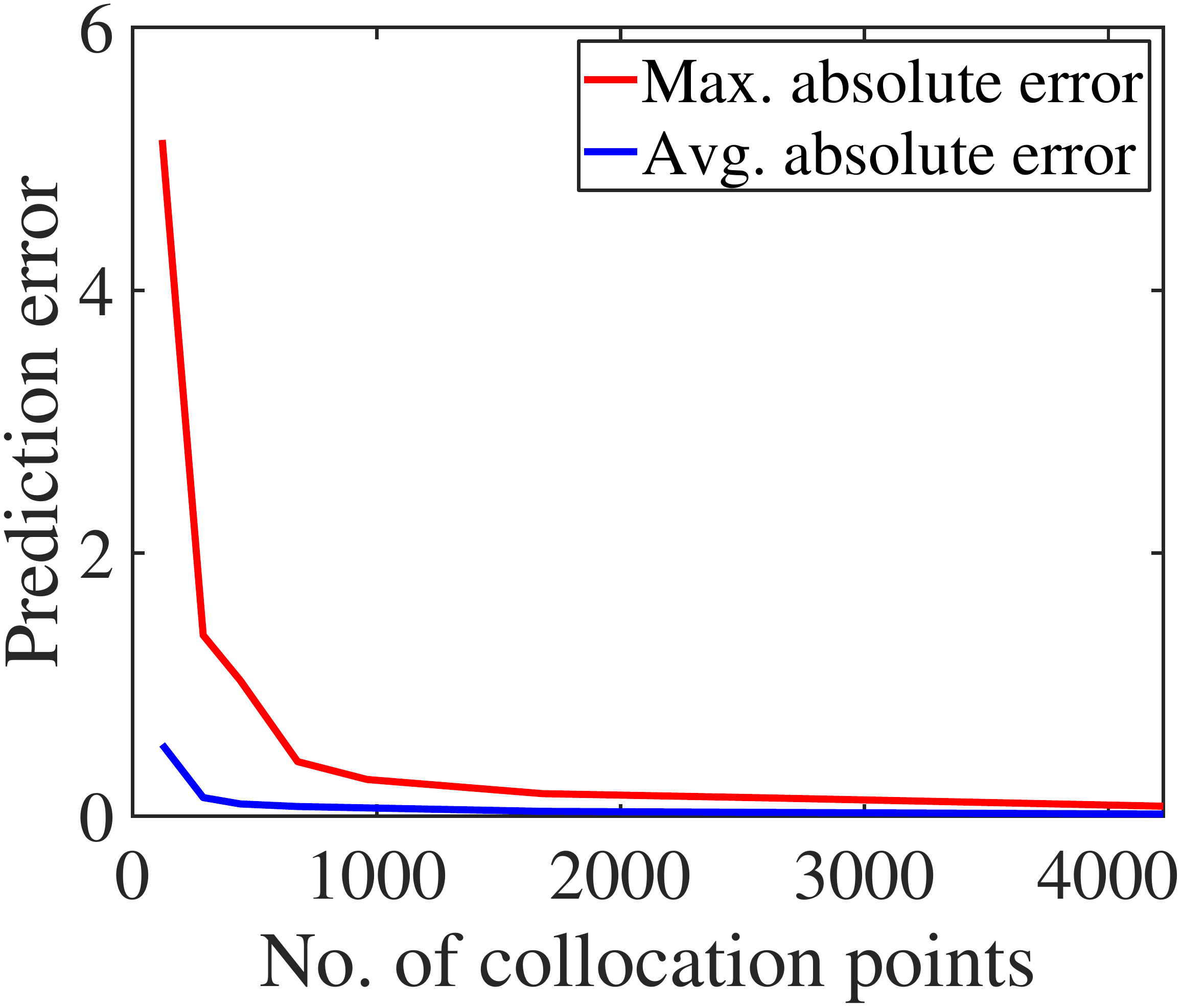}
    \caption{Variation of prediction error with total number of  collocation points for for the case of 1-D heat conduction example}
    \label{fig:convergence1}
\end{figure}

As mentioned earlier (\autoref{sec:pa}), stochastic projection method utilizes the neighbourhood information of a given point in evaluating the gradient. 
We studied the effect on number of neighborhood point on the proposed approach and the results obtained are shown in  \autoref{Table_heatcond_1}. For this study, the number of collocation points is fixed at 2601. The results indicate that the stochastic projection method obtains the gradients accurately with fewer neighboring points. This can be justified as the inclusion of relatively farther neighboring points in the formulation causes the local variation of the function to diminish. In addition, we also investigated the influence of the the number of collocation points on the performance of the SP-PINN, and the same is illustrated in \autoref{fig:convergence1}. 
As expected, performance of the proposed SP-PINN improves with increase in the number of collocation points. Another study on the effect of activation functions on performance of PINN is presented in the \autoref{Table_heatcond_2}. From the results, it can be infered that while SP-PINN with tanh and ELU activation functions yield accurate predictions, ReLU predicts the solution with significant prediction error ($4.71\%$). The study is carried out by keeping the number of neighboring points ($4$) and the number of collocation points ($2601$) constants. Here, we note that using ReLU activation with PINN for the heat conduction problem is not feasible with AD-PINN because of the differentiability requirement. SP-PINN, on the other hand, can achieve reasonable accuracy regardless of the activation functions.
\begin{table}[ht!]
    \centering
    \caption{Variation of prediction error with number of neighbouring collation points for the case of 1-D heat conduction example}
    \label{Table_heatcond_1}
\begin{tabular}{lccc} 
\hline
\textbf{Number of neighbours points} & \textbf{Max. absolute  error} & \textbf{Average error}\\\hline
4 &	 0.1241  &	0.01328 -\\
12 &	0.2125   & 0.03656 \\
24 & 	0.3709 & 0.04063 \\ 
36	& 0.2780  & 0.05418\\
\hline 
\end{tabular}
\end{table}

\begin{figure}[!h]
    \centering
    \subfigure[]{
    \includegraphics[width=.3 \textwidth]{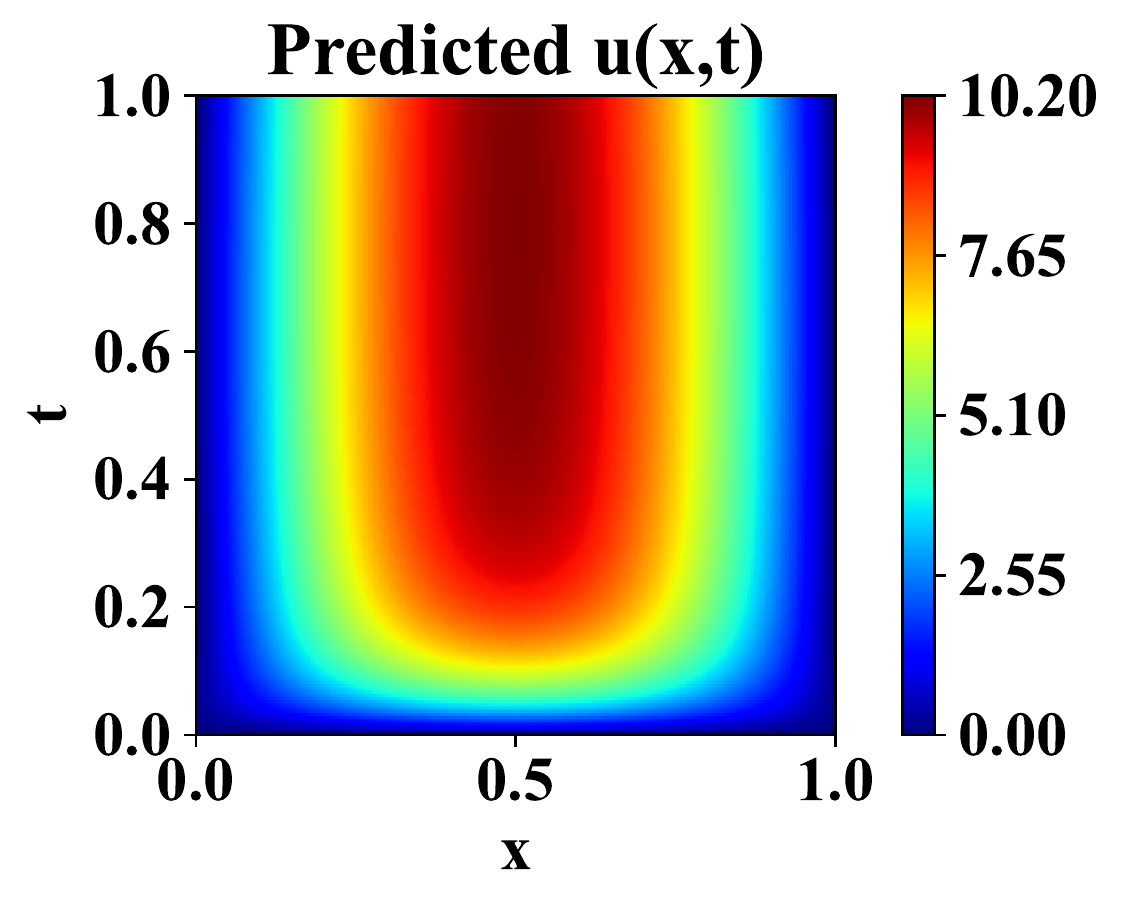}}
    \subfigure[]{
    \includegraphics[width=.3\textwidth]{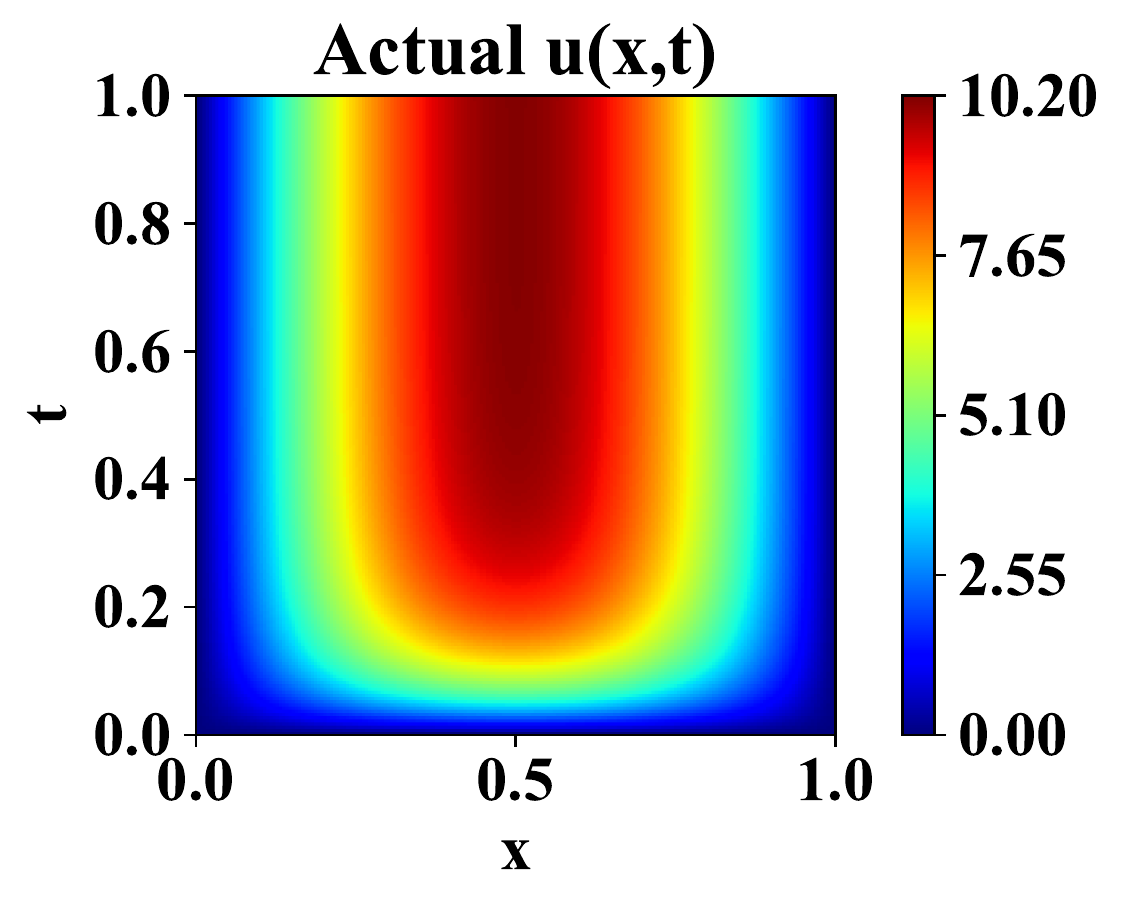}}
    \subfigure[]{
    \includegraphics[width=.3\textwidth]{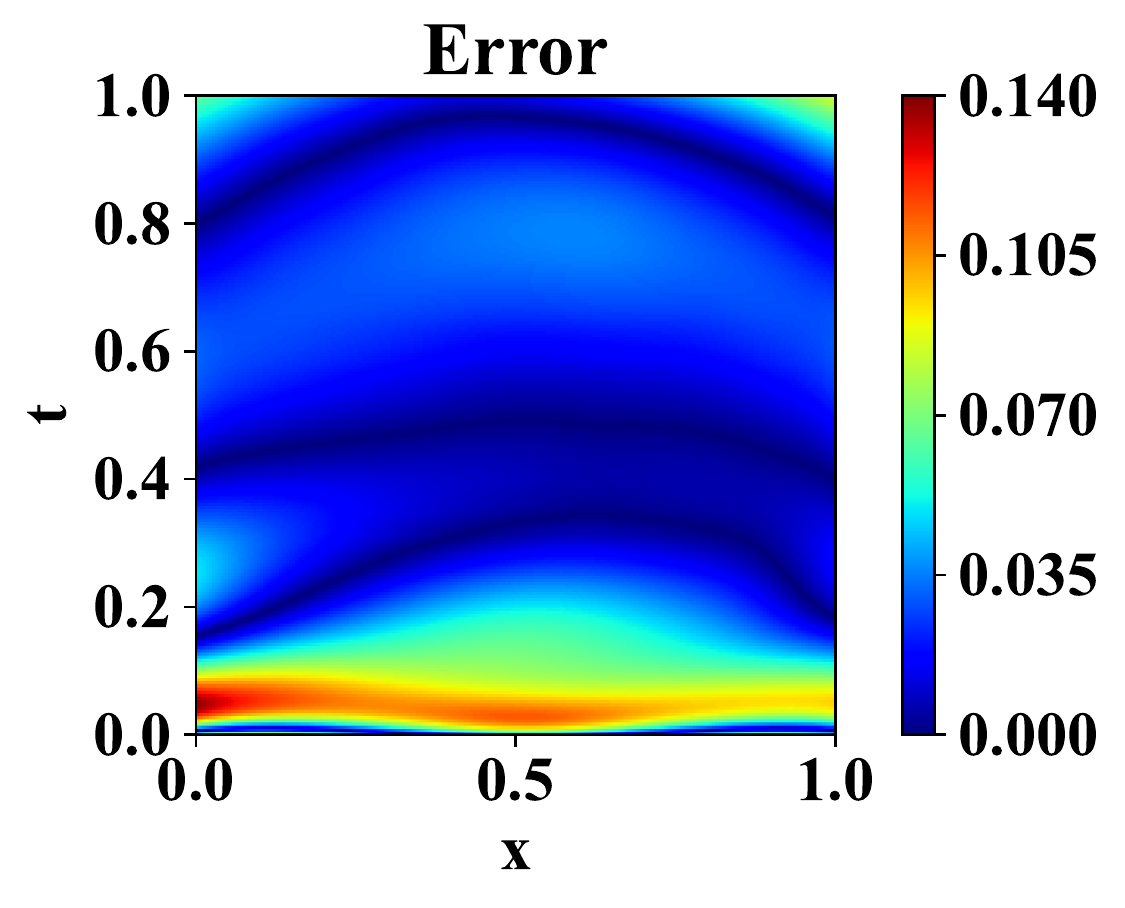}}
    \caption{Results of the 1-D heat conduction example with 2601 collocation points; (a) Solution predicted by the SP-PINN, (b) Actual results, (c) Absolute error of predicted solution with the actual solution }
    \label{fig:case11}
\end{figure}

\begin{figure}[!h]
    \centering
    \subfigure[]{
    \includegraphics[width=.3 \textwidth]{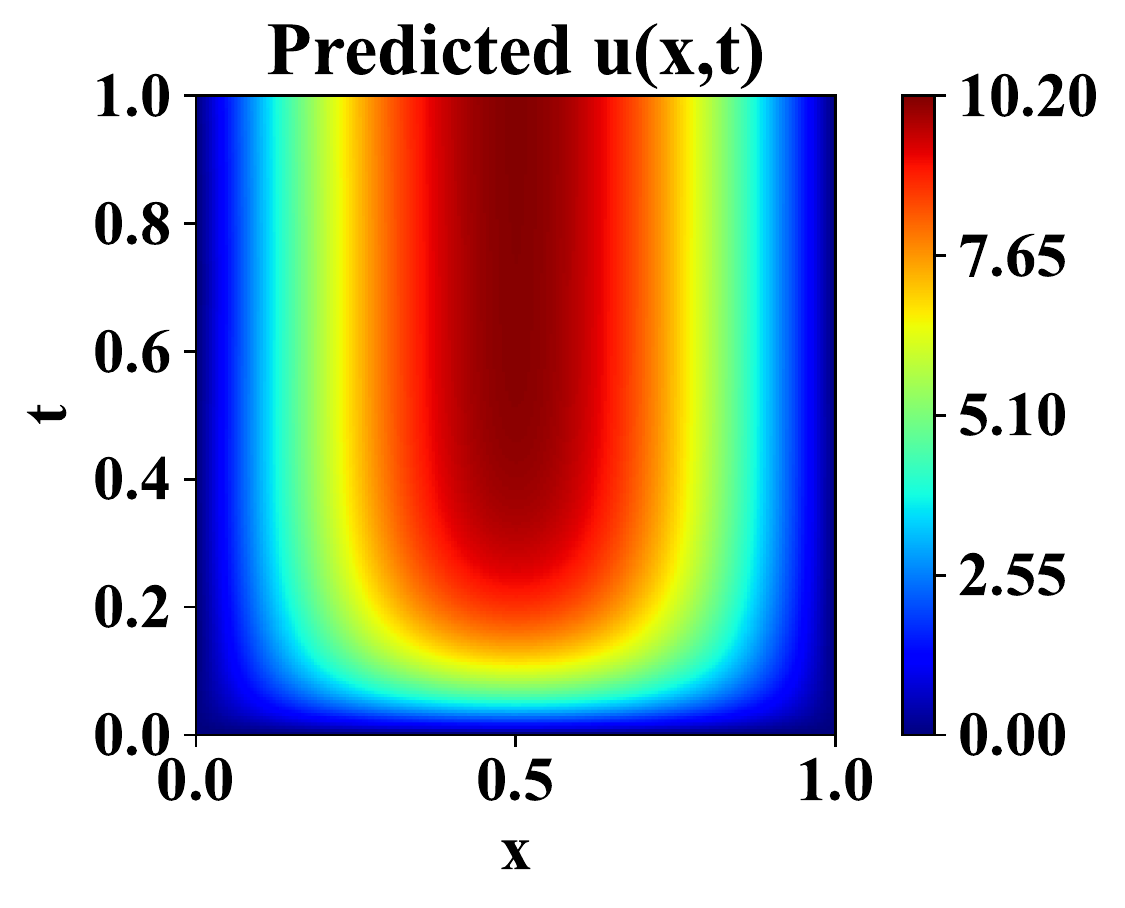}}
    \subfigure[]{
    \includegraphics[width=.3\textwidth]{Heatconduction_actual.pdf}}
    \subfigure[]{
    \includegraphics[width=.3\textwidth]{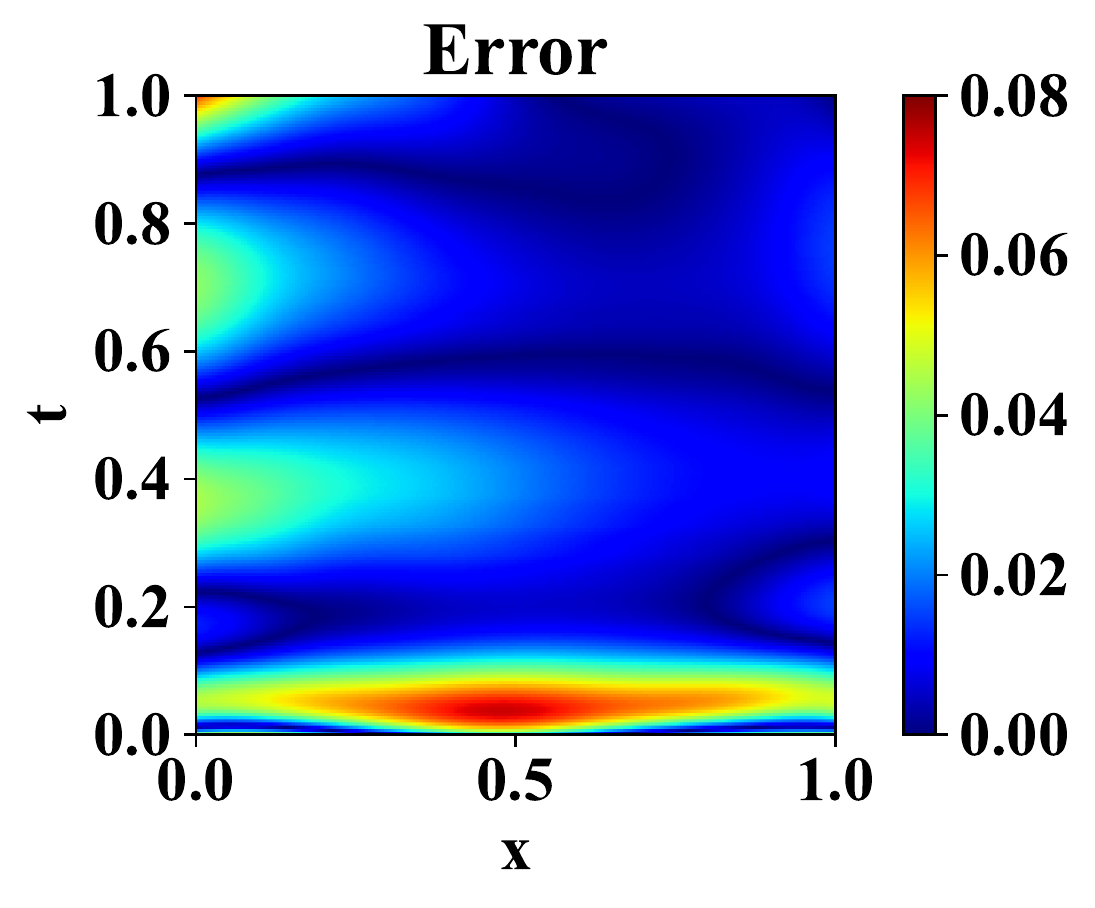}}
    \caption{Results of the 1-D heat conduction example with 4225 collocation points; (a) Solution predicted by the SP-PINN, (b) Actual results, (c) Absolute error of predicted solution with the actual solution }
    \label{fig:case12}
\end{figure}

\begin{table}[ht!]
    \centering
    \caption{Prediction error with architectures utilising activation functions ELU, Tanh and RelU for the case of 1-D heat conduction example}
    \label{Table_heatcond_2}
\begin{tabular}{lccc} 
\hline
\textbf{activation functions} & \textbf{Max. absolute  error} & \textbf{Average error}\\\hline
ELU	& 0.2227 & 0.02751\\
Tanh & 0.10621 & 0.02548\\
ReLU & 4.71098 & 1.5616\\
\hline 
\end{tabular}
\end{table}
Based on the case studies presented above, we selected ELU activation function and four neighboring points. The contou plot of the result are presented in Figs. \ref{fig:case11} and \ref{fig:case12}. While \autoref{fig:case11} corresponds to SP-PINN with 2601 collocation points, \autoref{fig:case12} corresponds to SP-PINN trained with 4225 collocation points. For both the cases, the contours matches almost exactly with the ground truth with maximum prediction errors of 0.138 and 0.0746, respectively.

\subsection{Viscous Burger's Equation}
\begin{figure}[!h]
    \centering
    \includegraphics[width=.4 \textwidth]{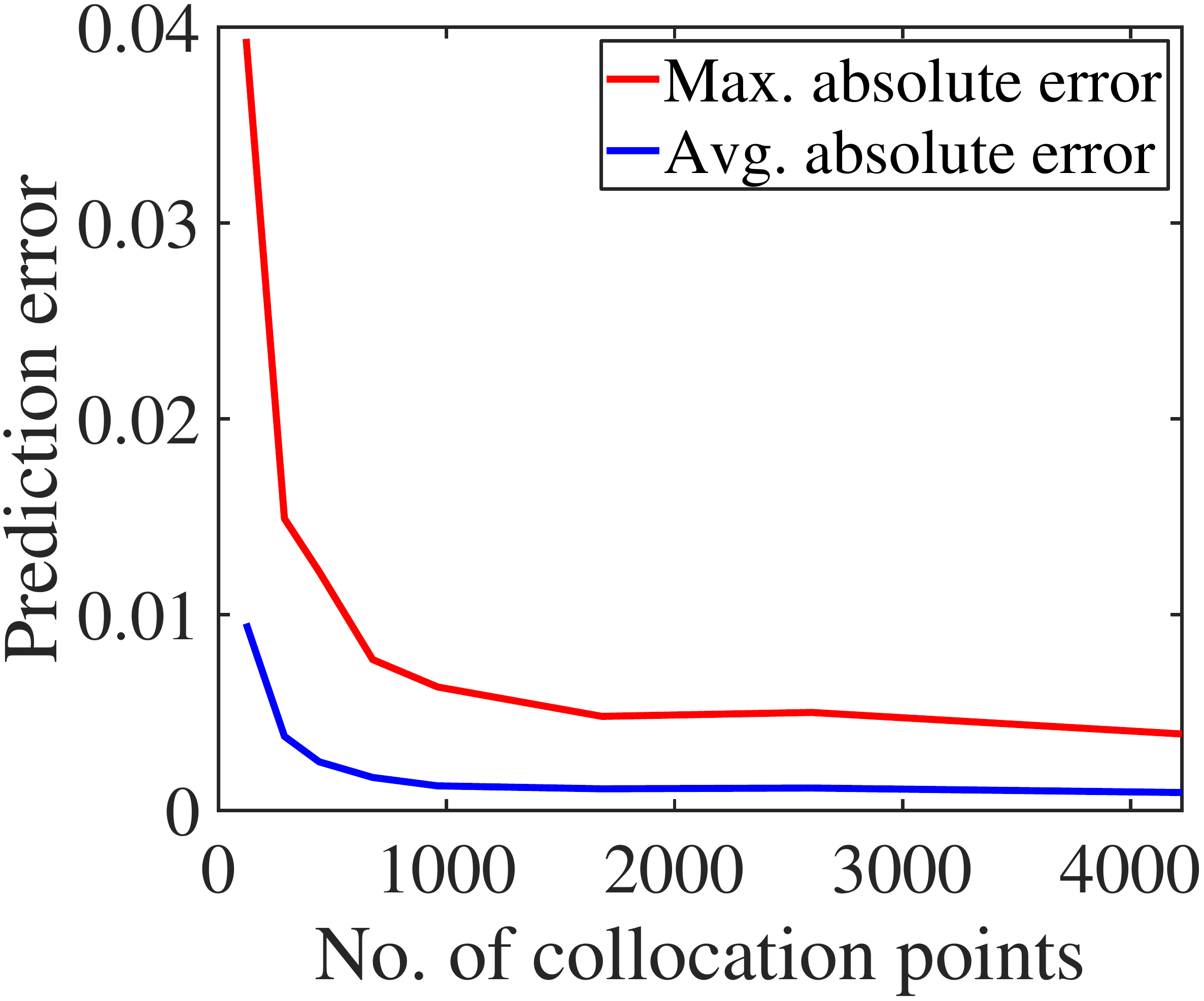}
    \caption{Variation of prediction error with number of neighbouring collocation points for the case of 1-D Burger example}
    \label{fig:convergence2}
\end{figure}

To further illustrate the efficacy of the proposed framework, we consider a nonlinear advection equation widely used in fluid dynamics and gas dynamics to describe shock wave propagation. Through this example, we validate the performance of the SP-PINN in solving nonlinear time-dependent problems. The governing PDE of the system is expressed as follows:
\begin{equation}\label{burgers equation}
 u_t + u u_x - u_{xx} = 0 ,\hspace{2em}   x\in [0,l],\hspace{2em}   t\in[0,T]
\end{equation}
with boundary condition: $u(0,t)=u(l,t)=0$ and initial condition: $u(x,0)= sin(\pi{x}/l)$. Here the length parameter and time period are set such that; $l=1$ and $T=1$. For predicting the solution, we employ the identical architecture employed in the first example along with the L-BFGS optimizer keeping the learning rate, $l_r=0.1$. The generated points, $N_b=400$ at $x=0$, $x=1$ and $T=0$ are utilised to enforce the boundary and initial conditions. Similar to the previous example, the influence of the number of collocation points, number of neighborhood points, and activation functions are studied. The results are provided in the \autoref{table_ex21}, \autoref{table_ex22} and \autoref{fig:convergence2}. \autoref{table_ex21} shows the variation of the prediction error with the number of neighboring points, which reinforces the fact that with less number of neighborhood points yields  a more accurate result. Moreover, according to \autoref{fig:convergence2}, the convergence of the results with the number of collocation points is seen in this case as well.
\begin{table}[ht!]
    \centering
    \caption{Variation of prediction error with number of neighbouring collocation points for the case of 1-D Burger example}
    \label{table_ex21}
\begin{tabular}{lccc} 
\hline
\textbf{Number of neighbours points} & \textbf{Max. absolute  error} & \textbf{Average error}\\\hline
4 &	0.007592  & 0.00126 \\
12 & 0.011054  & 0.002026 \\
24 & 0.0106553  & 0.002680 \\
36 & 0.0132238  & 0.003235\\
\hline 
\end{tabular}
\end{table}

\begin{table}[ht!]
    \centering
    \caption{Prediction error with architectures utilising activation functions ELU, Tanh and RelU for the case of 1-D Burger example}
    \label{table_ex22}
\begin{tabular}{lccc} 
\hline
\textbf{Activation functions} & \textbf{Max. absolute  error} & \textbf{Average error}\\\hline
ELU &	0.0105647 &	0.001538\\
Tanh &	0.00964 &	0.001186 \\
ReLU &	0.06807 &	0.01715 \\
\hline 
\end{tabular}
\end{table}

\begin{figure}[!h]
    \centering
    \subfigure[]{
    \includegraphics[width=.3 \textwidth]{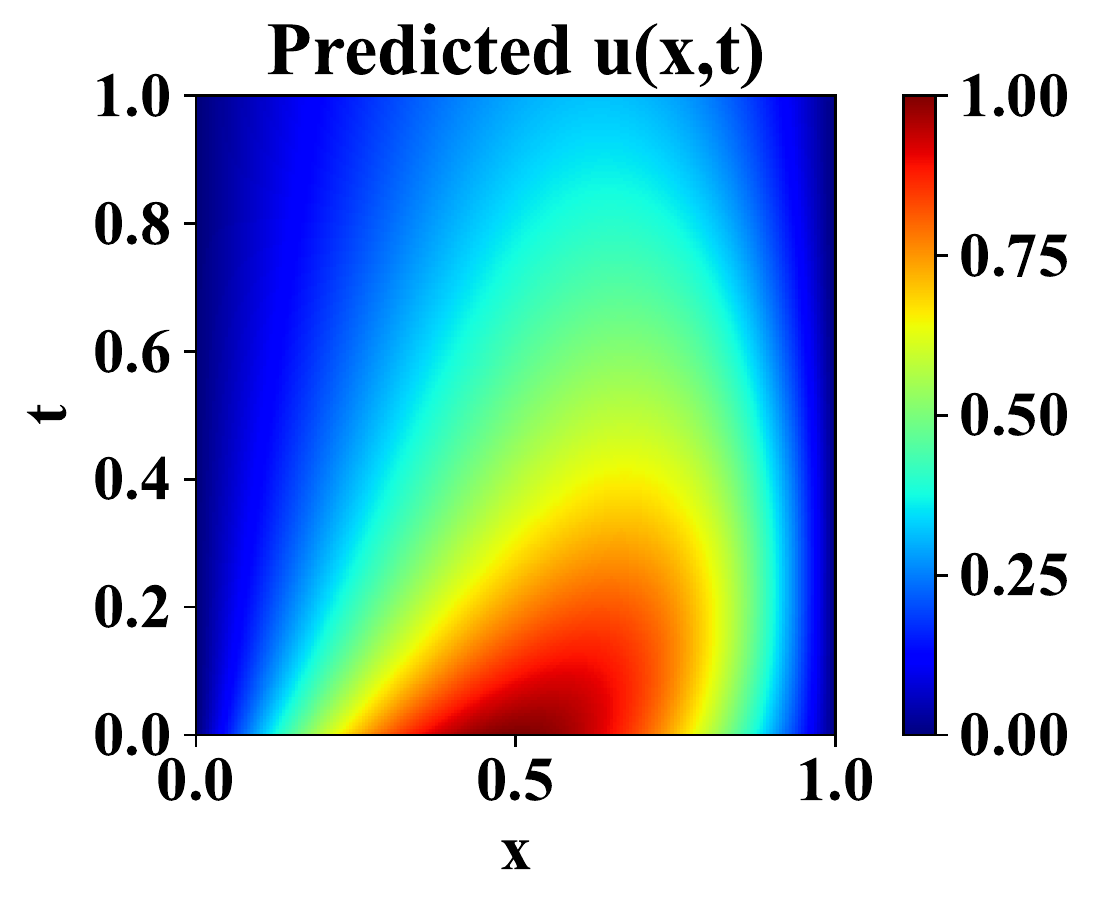}}
    \subfigure[]{
    \includegraphics[width=.3\textwidth]{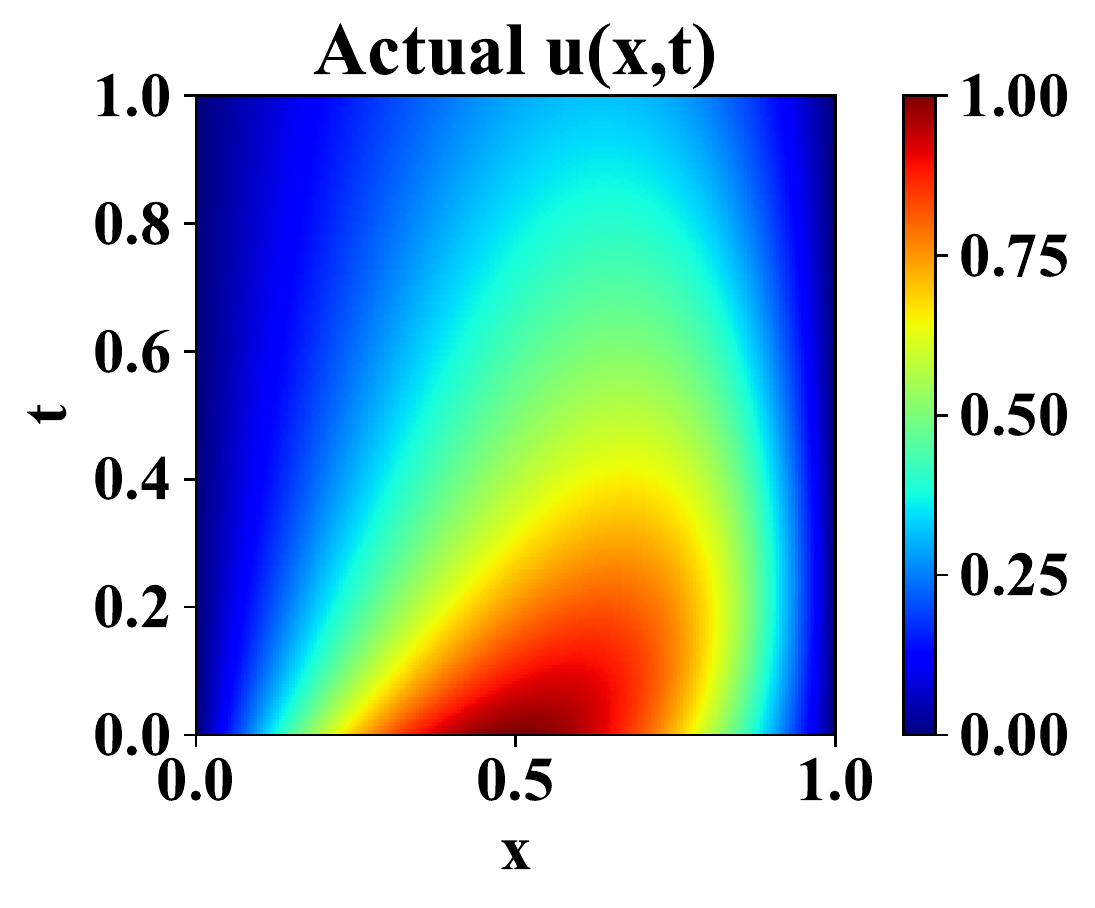}}
    \subfigure[]{
    \includegraphics[width=.3\textwidth]{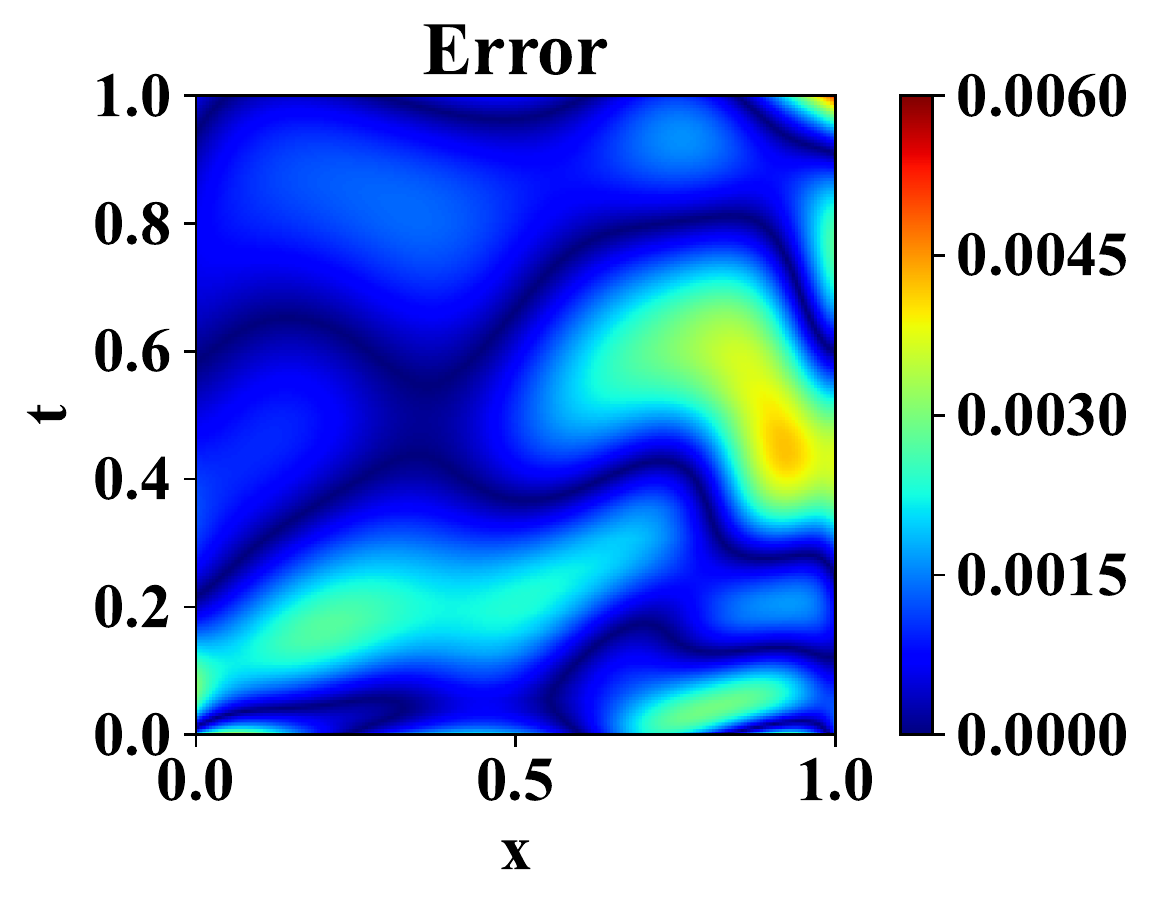}}
    \caption{Results of the 1-D Burger example with 2601 collocation points; (a) Solution predicted by the SP-PINN, (b) Actual results, (c) Absolute error of predicted solution with the actual solution}
    \label{fig:case21}
\end{figure}

Figs. \ref{fig:case21} and \ref{fig:case12} depict the pictorial depiction of results for 2601 and 4225 collocation points, respectively. The results obtained using SP-PINN shows excellent agreement with the actual solution, with a prediction error of less than $0.006$.
\begin{figure}[!h]
    \centering
    \subfigure[]{
    \includegraphics[width=.3 \textwidth]{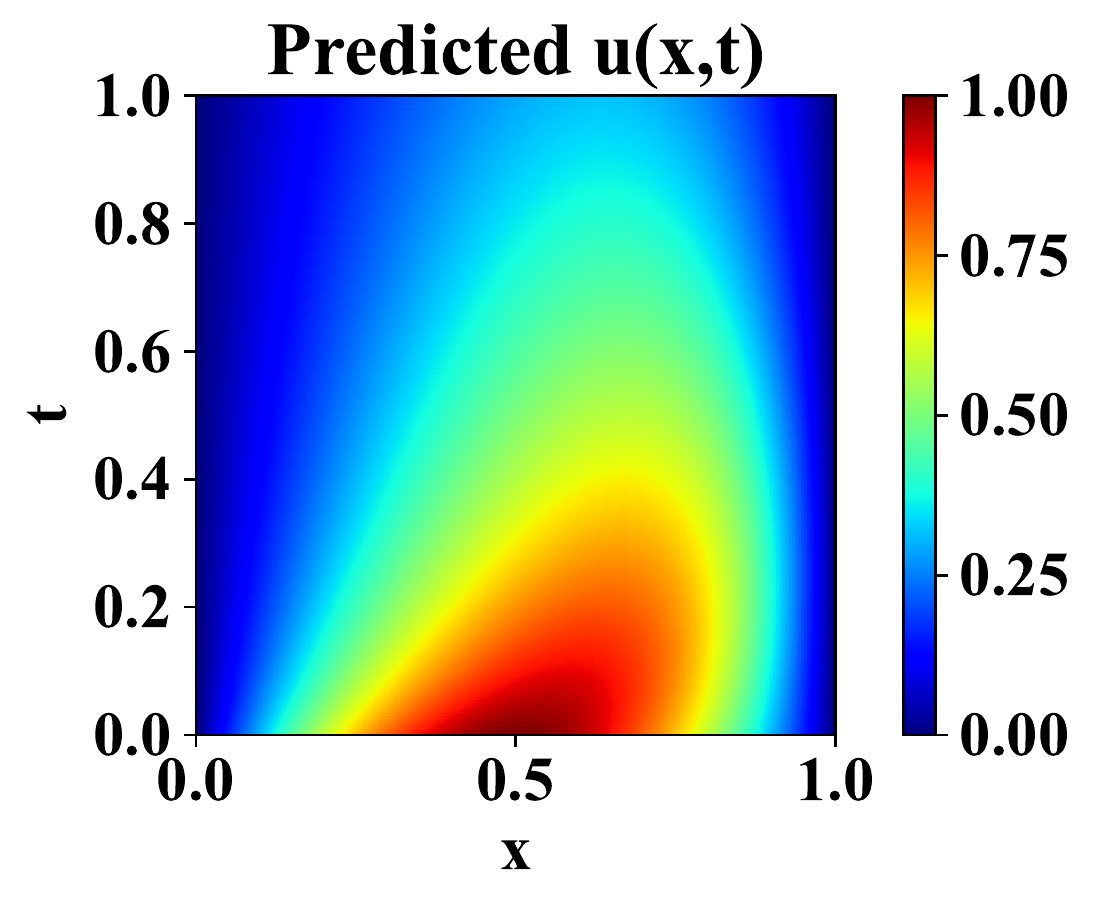}}
    \subfigure[]{
    \includegraphics[width=.3\textwidth]{Burger_actual.pdf}}
    \subfigure[]{
    \includegraphics[width=.3\textwidth]{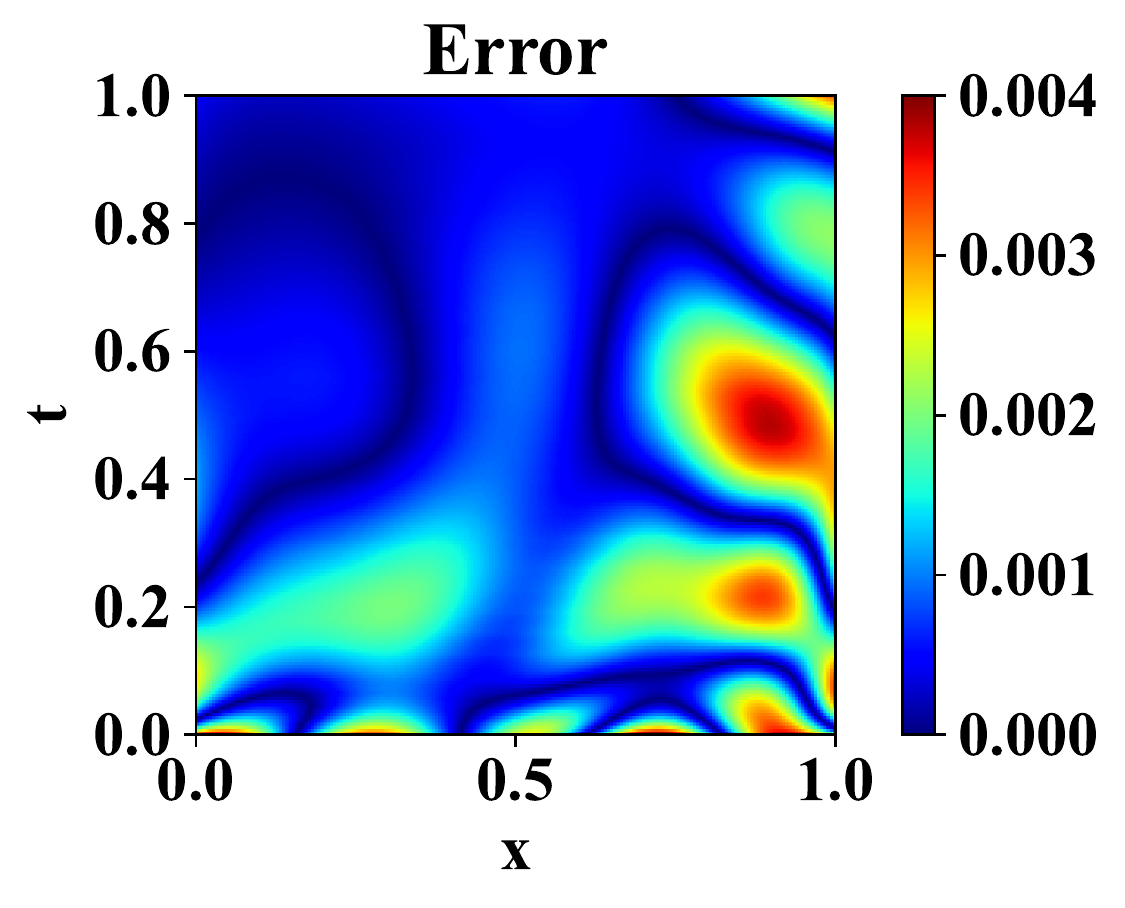}}
    \caption{Results of the 1-D Burger example with 4225 collocation points; (a) Solution predicted by the SP-PINN, (b) Actual results, (c) Absolute error of predicted solution with the actual solution}
    \label{fig:case22}
\end{figure}

\subsection{Poisson's equation}
\begin{figure}[!h]
    \centering
    \includegraphics[width=.5 \textwidth]{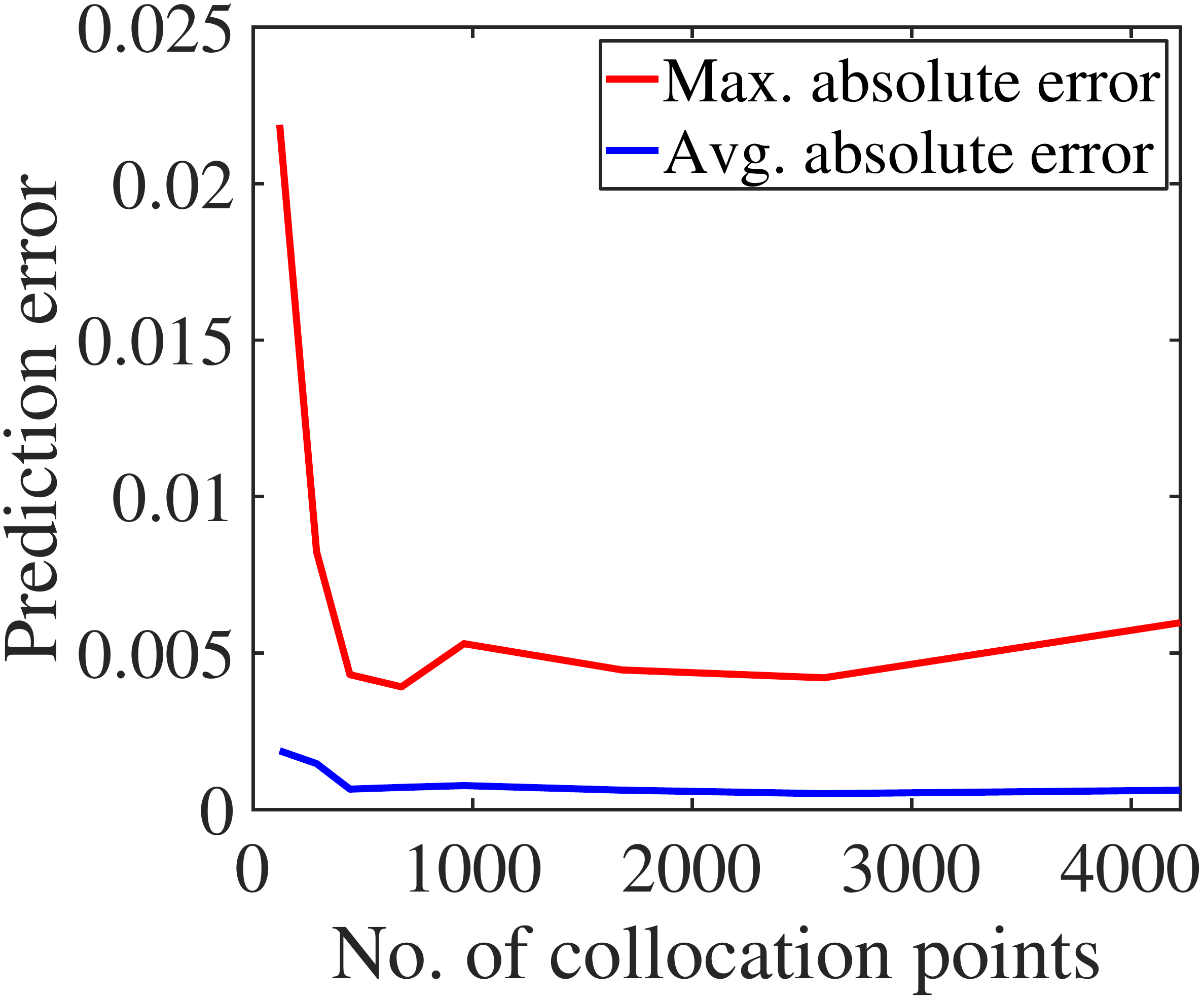}
    \caption{Variation of prediction error with total number of  collocation points for for the case of 2-D Poisson's equation example}
    \label{fig:convergence3}
\end{figure}

As the third example, we elucidate the application of the proposed SP-PINN in solving the Poisson's equation. This equation describes the governing PDE for the case of steady state heat conduction with source function and is mathematically represented as:
\begin{equation}\label{Poissons equation1}
 u_{xx} + u_{yy} = f(x,y) ,\hspace{2em}   x\in [-1,1],\hspace{2em}  y\in[-1,1]
\end{equation}
We consider homogeneous Poisson's equation and hence, $f(x,y)=0$. A two-dimensional domain is described by $x$ and $y$ coordinates. Since the problem is time-independent, other than the collocation points, only randomly generated boundary points are used for training the network. The network architecture, optimizer, and learning rate are considered to be the same as in the previous two examples. The studies on the impact of the number of neighboring points and activation functions are presented in the Tables \ref{table_ex31} and \ref{table_ex32}, while the results depicting the convergence of prediction error with the number of collocation points are provided in the \autoref{fig:convergence3}. It is apparent from the results that, contrary to the first and second numerical examples, the results listed in the \autoref{table_ex31} show no substantial increment in the prediction error with the number of neighborhood points with eight neighboring points yielding slightly better results. On the other hand, \autoref{fig:convergence3} provides a clear indication of the convergence of the prediction error with a number of collocation points.
\begin{table}[ht!]
    \centering
    \caption{Variation of prediction error with number of neighbouring collocation points for th case of 2-D Poisson's equation example}
    \label{table_ex31}
\begin{tabular}{lccc} 
\hline
\textbf{Number of neighbours points} & \textbf{Max. absolute  error}  & \textbf{Average error}\\\hline
8  &	0.001235  & 0.000248\\ 
20 &	0.001170  & 0.000263\\
28 &	0.0013006  & 0.000273\\
36 &	0.0015151  & 0.0003038\\
\hline 
\end{tabular}
\end{table}

\begin{table}[ht!]
    \centering
    \caption{Prediction error with architectures utilising activation functions ELU, Tanh and RelU for the case of 2-D Poisson's equation example}
    \label{table_ex32}
\begin{tabular}{lccc} 
\hline
\textbf{Activation functions} & \textbf{Max. absolute  error}  & \textbf{Average error}\\\hline
ELU	& 0.00353	& 0.00048\\
Tanh &	0.00355 & 0.000699 \\
ReLU &	0.12952 &	0.02604\\
\hline 
\end{tabular}
\end{table}

\begin{figure}[!h]
    \centering
    \subfigure[]{
    \includegraphics[width=.3 \textwidth]{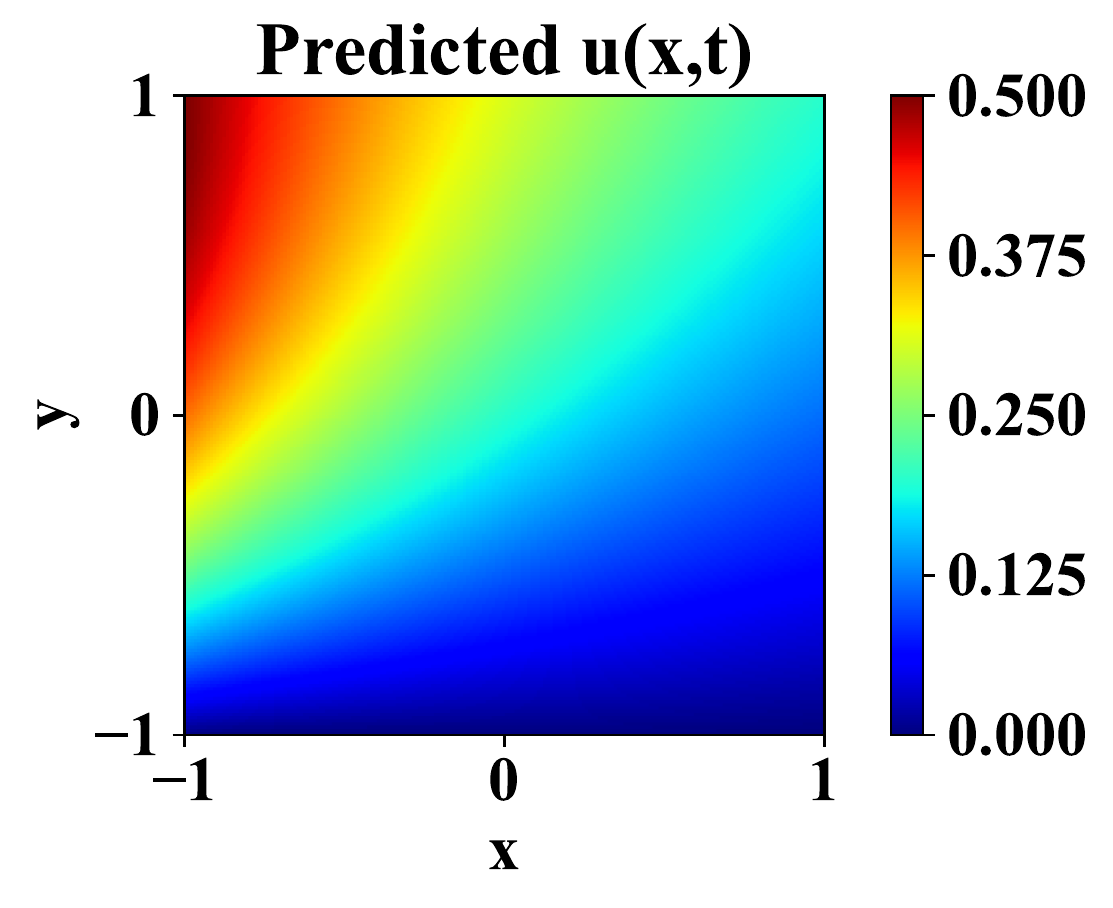}}
    \subfigure[]{
    \includegraphics[width=.3\textwidth]{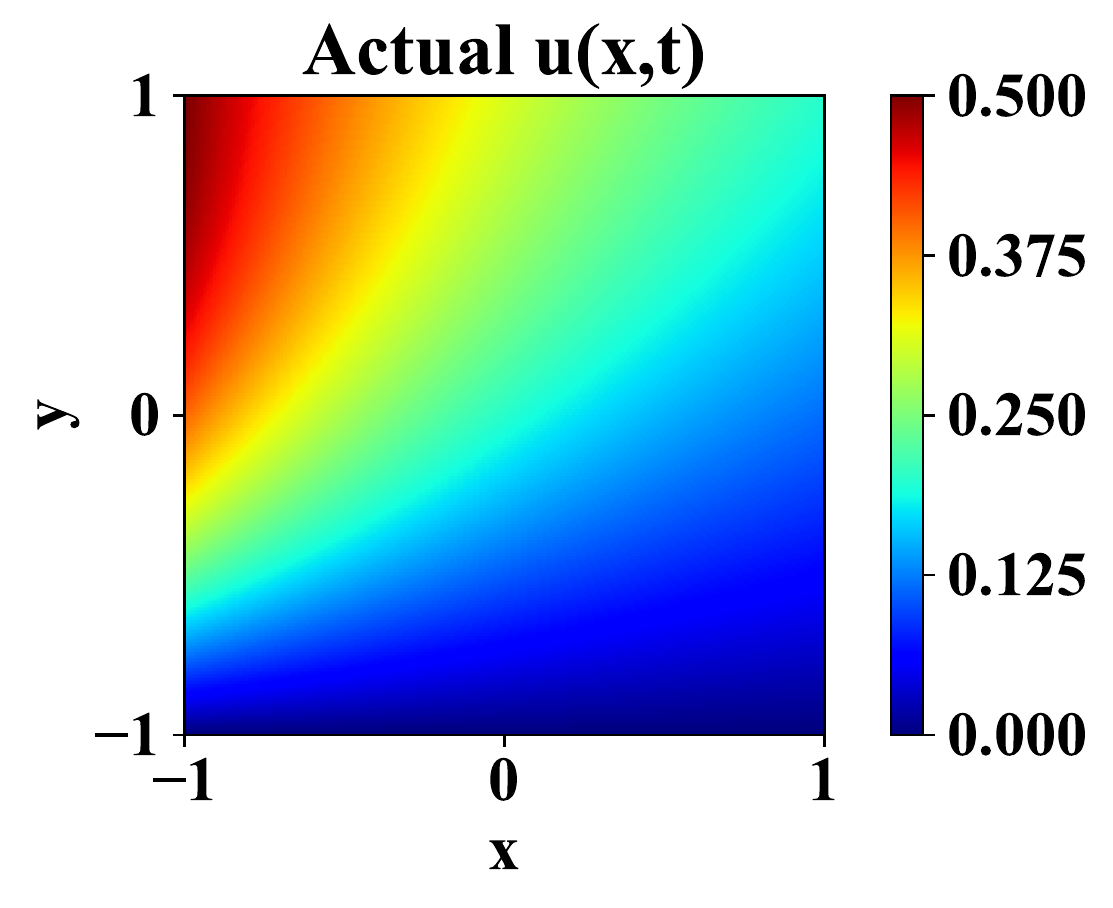}}
    \subfigure[]{
    \includegraphics[width=.3\textwidth]{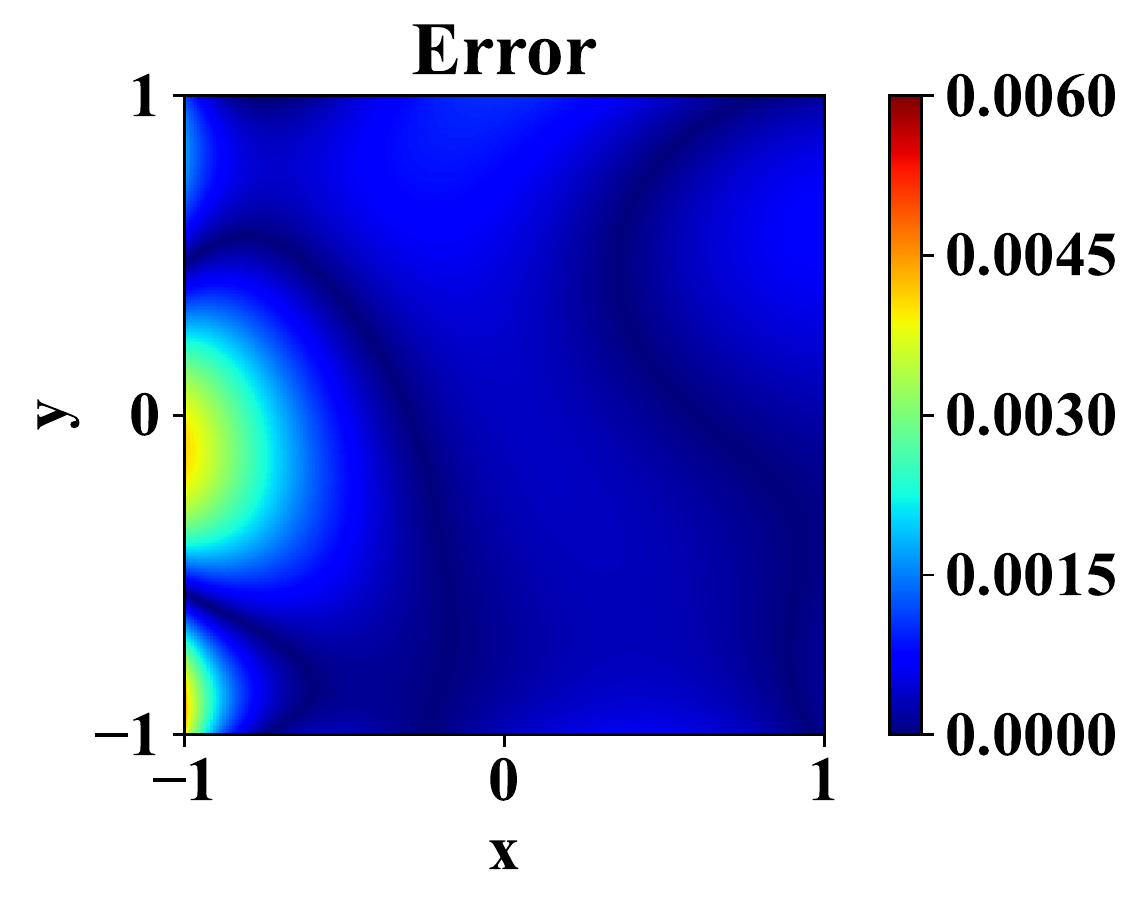}}
    \caption{Results of the 2-D Poisson's equation with 2601 collocation points; (a) Solution predicted by the SP-PINN, (b) Actual results, (c) Absolute error of predicted solution with the actual solution}
    \label{fig:case31}
\end{figure}

\begin{figure}[!h]
    \centering
    \subfigure[]{
    \includegraphics[width=.3 \textwidth]{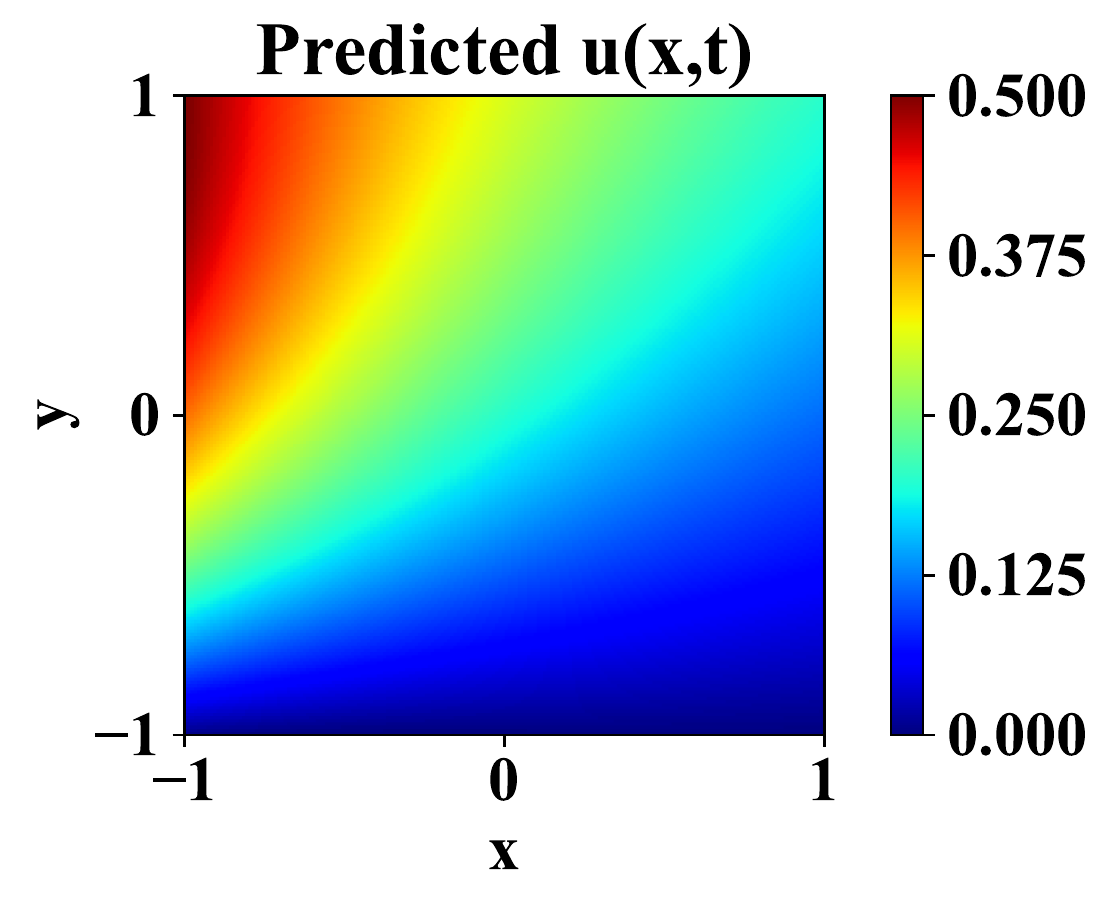}}
    \subfigure[]{
    \includegraphics[width=.3\textwidth]{Poissons1_actual.pdf}}
    \subfigure[]{
    \includegraphics[width=.3\textwidth]{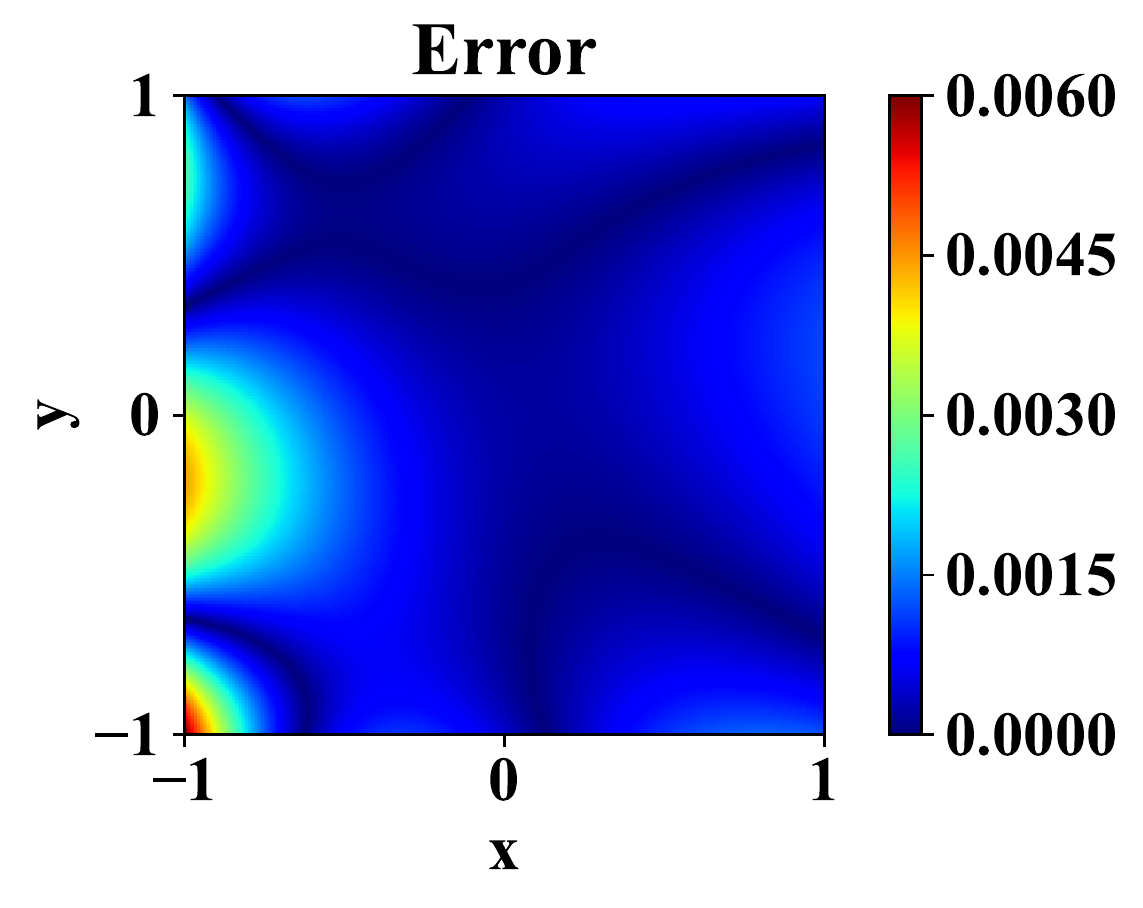}}
    \caption{Results of the 2-D Poisson's equation with 4225 collocation points; (a) Solution predicted by the SP-PINN, (b) Actual results, (c) Absolute error of predicted solution with the actual solution}
    \label{fig:case32}
\end{figure}

The contours obtained using SP-PINN are shown in Figs. \ref {fig:case31} (2601 collocation points) and \ref{fig:case32} (4335 collocation points). 
The plots apparently showcase the close agreement of the SP-PINN predicted results with the  actual solution with prediction error less than $0.006$. 
\subsection{Comparison AD-PINN and SP-PINN for problems involving irregular geometry and non-smooth solution}\label{comp1}
Till now, we focused on application of SP-PINN to problems defined on regular domain. However, on regular domains, AD-PINN already yields accurate results. In this section, we present some standard benchmark problems involving irregular domain and non-smooth solution. The objective here is to compare the performance of the proposed SP-PINN and vanilla PINN (aka AD-PINN). 
\subsubsection{Non-homogeneous Poisson's equation}\label{subsubsec:n_poisson}
\begin{figure}[!h]
    \centering
    \subfigure[]{
    \includegraphics[width=.3\textwidth]{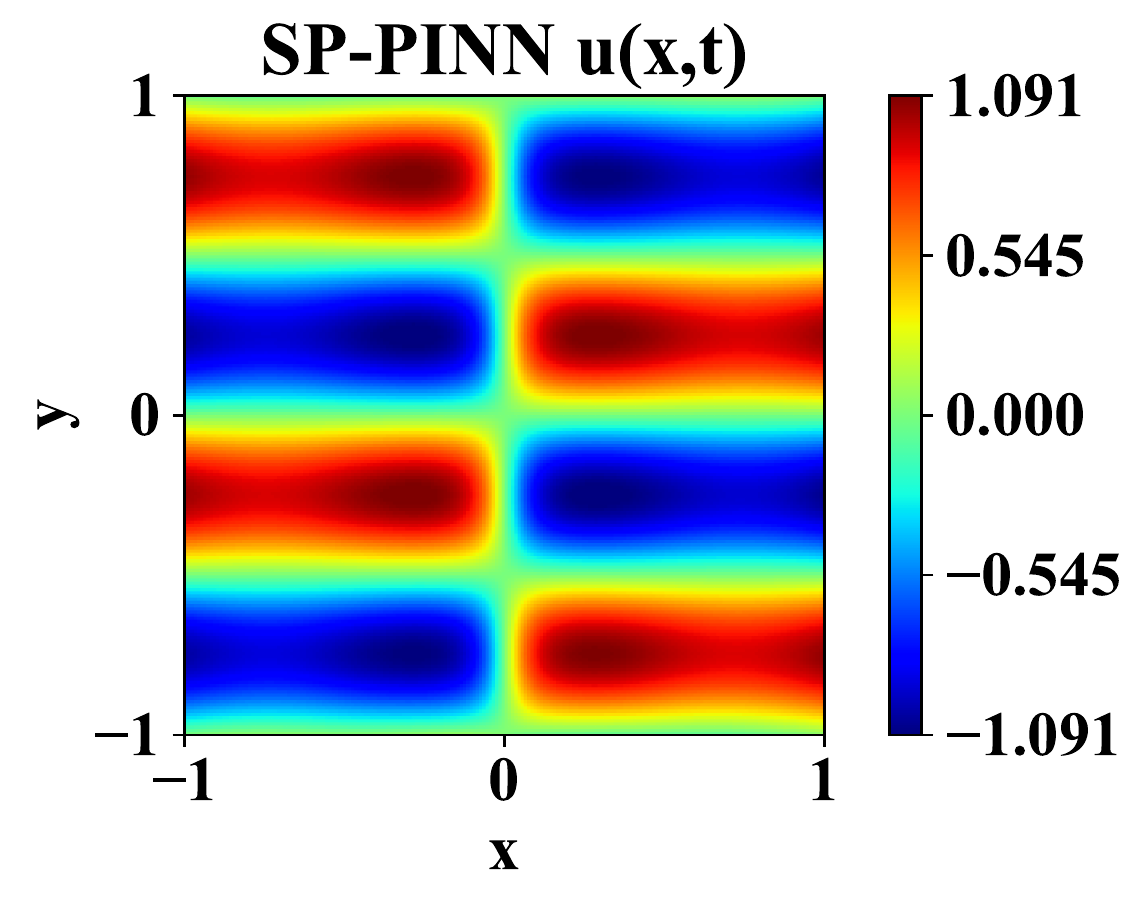}}
    \subfigure[]{
    \includegraphics[width=.3\textwidth]{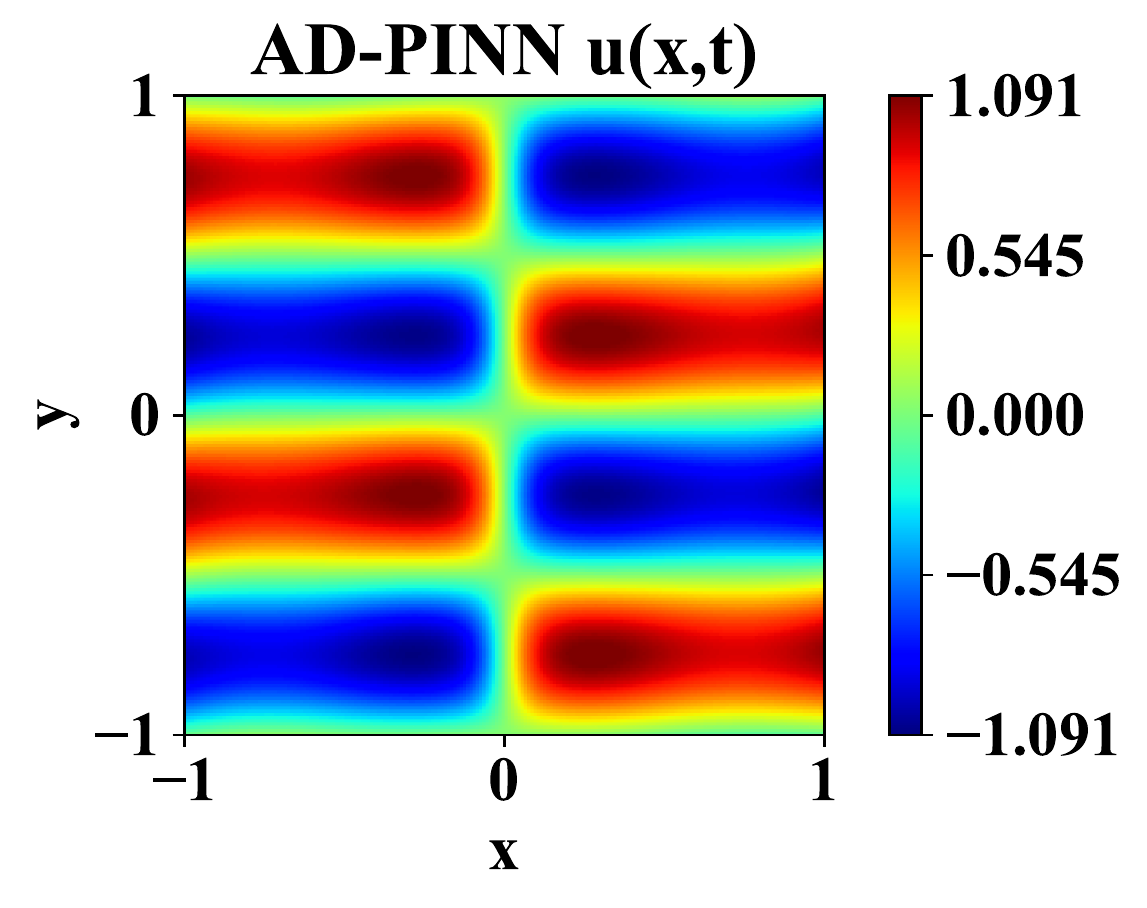}}
    \subfigure[]{
    \includegraphics[width=.3\textwidth]{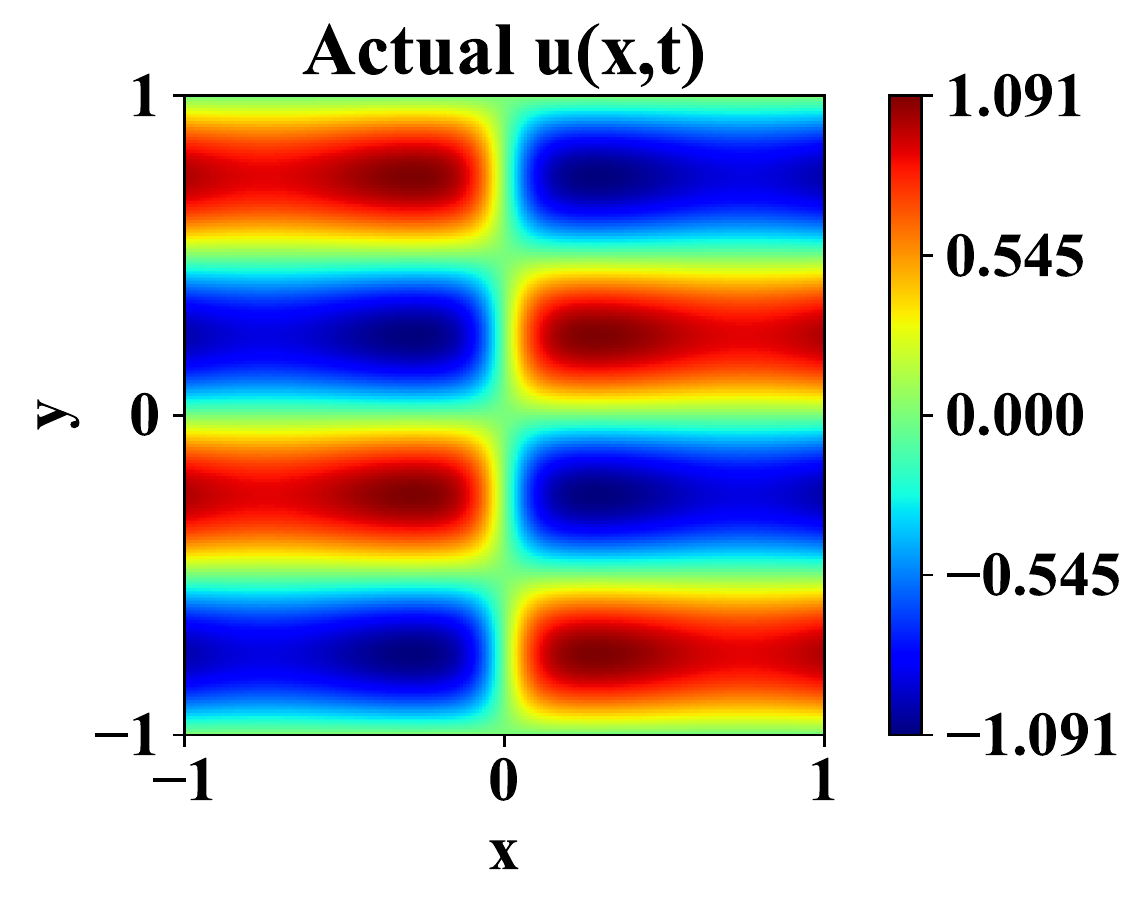}}
    \caption{Results of the non-homogeneous Poisson’s example with non zero source function with 2601 collocation points; (a) Solution predicted by the SP-PINN, (b) AD-PINN, (c) Actual solution}
    \label{fig:case41}
\end{figure}

\begin{figure}[!h]
    \centering
    \subfigure[]{
    \includegraphics[width=.3\textwidth]{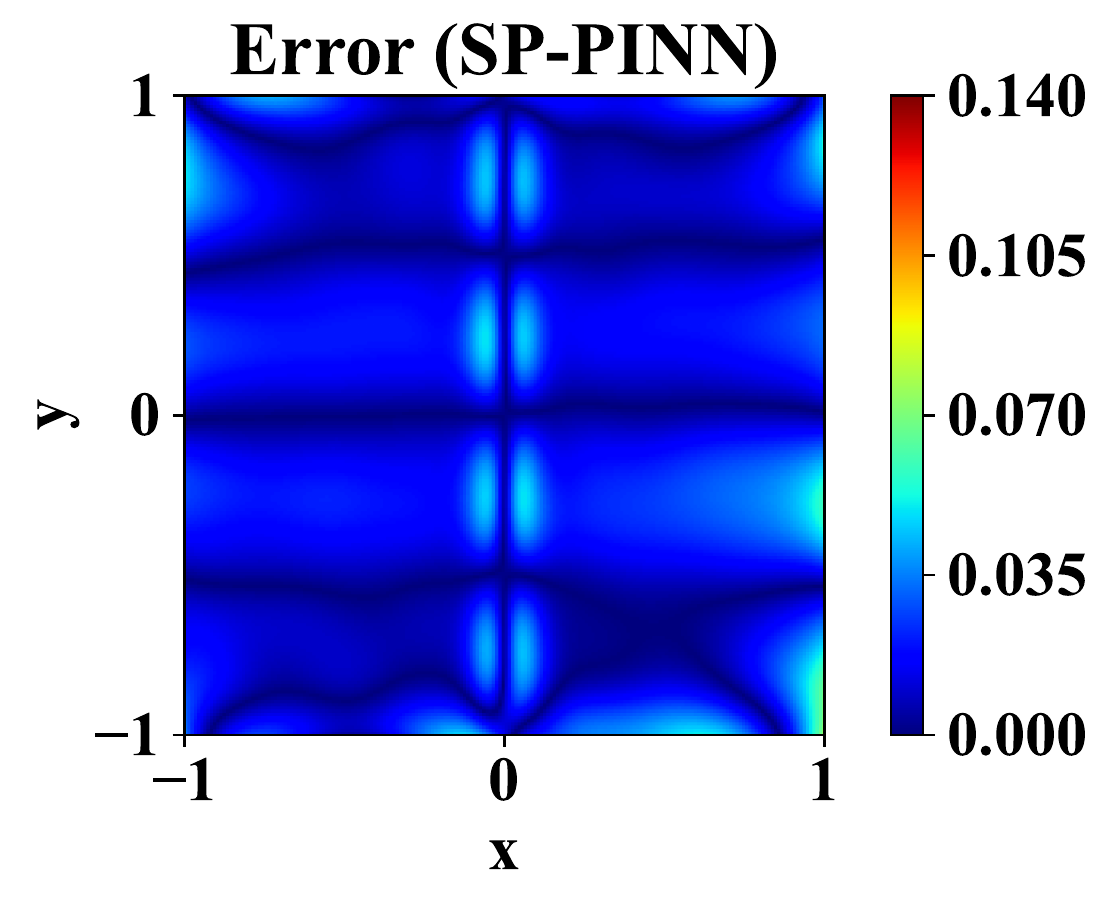}}
    \subfigure[]{
    \includegraphics[width=.3\textwidth]{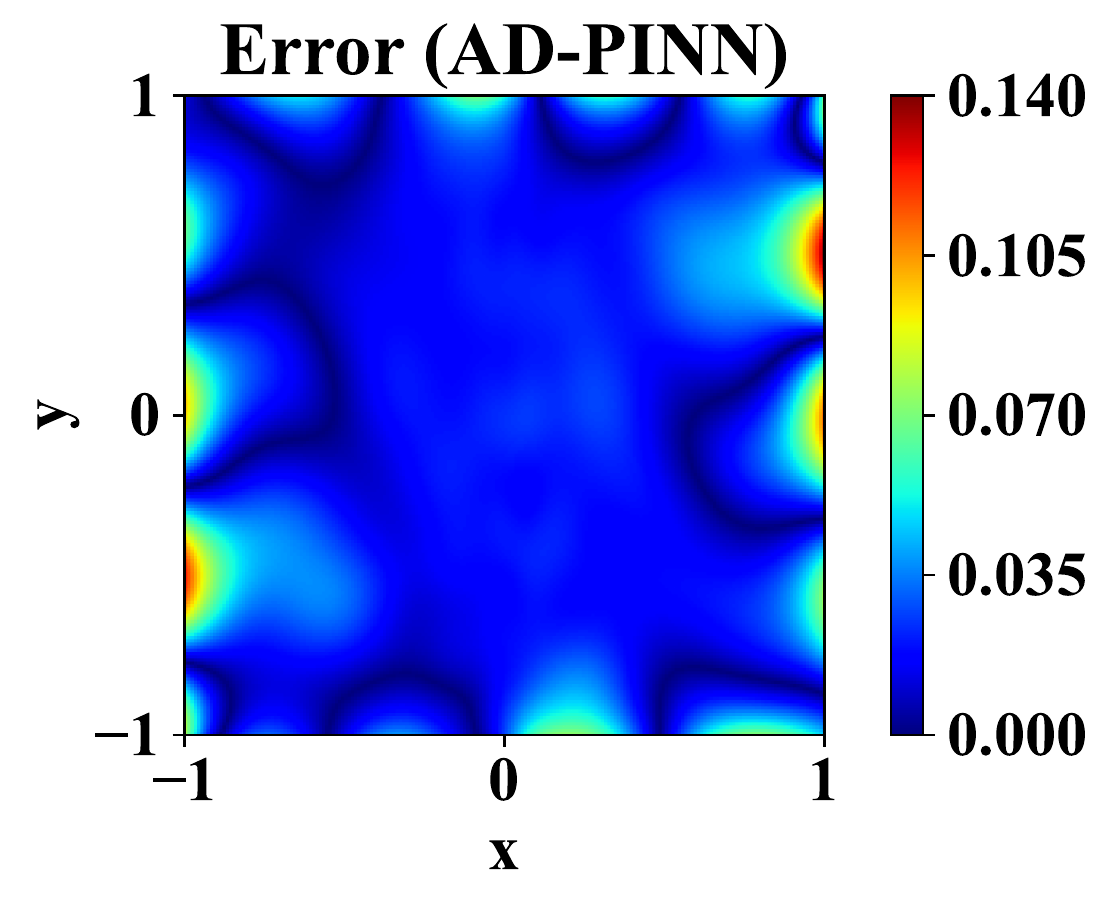}}
    \caption{Prediction error of the non-homogeneous Poisson’s example with non zero source function with 2601 collocation points; (a) The absolute error of SP-PINN. (b) The absolute error of AD-PINN}
    \label{fig:case42}
\end{figure}

We start by considering the non-homogeneous Poisson's equation. The governing equation in this case remains same as \autoref{Poissons equation1}; however, we now have 
$f(x,y) = \sin(2\pi{y})[{20}\tanh({10}x)({10}\tanh^2(10{x}) - 10) - {2\pi^2}\sin(2\pi{x}/{5}) - {4\pi^2}\sin(2\pi{y})(\tanh({10}{x})+\sin(2\pi{x}/10)]$
For a fair comparison, we utilize same number of collocation points for training both SP-PINN and AD-PINN. The architecture of the neural networks employed in both the cases consist of 3 hidden layers with 40, 80 and 40 neurons, and \texttt{tanh} activation function is used. We use a learning rate of $l_r=0.1$. We emphasize here that only the gradient computation differs between the AD-PINN and SP-PINN models. The results of the predictions of the models are illustrated here. We studied two cases; in the first case the number of collocation points used for training the model are 2601, whereas in the second case 4225 collocation points are used. The contours of the response are presented in Figs. \ref{fig:case41} and \ref{fig:case42}. Prediction error for SP-PINN and AD-PINN are also listed in the \autoref{Table_Poissons22}. 
The results indicate that as the solution over the domain becomes complex, the SP-PINN outperforms the AD-PINN, even with less number of collocation points. 

\begin{figure}[!h]
    \centering
    \subfigure[]{
    \includegraphics[width=.3\textwidth]{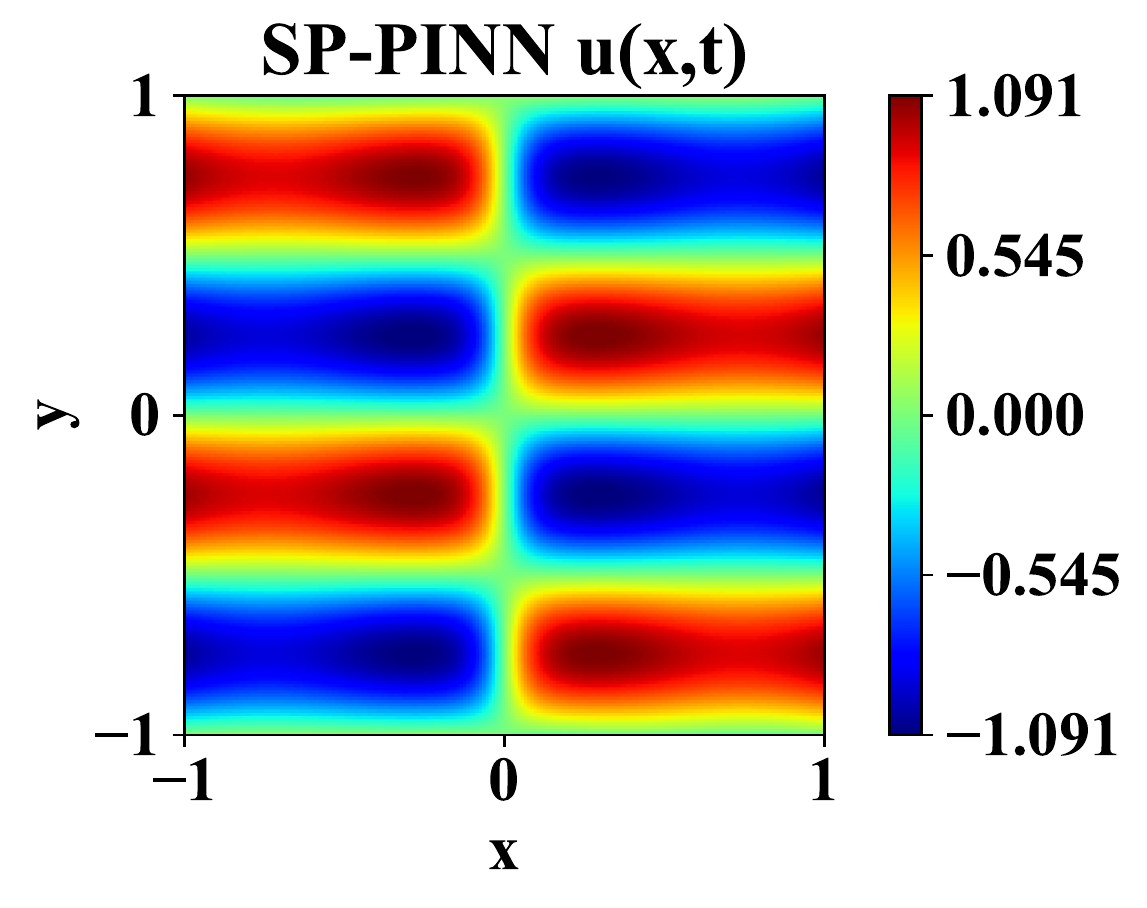}}
    \subfigure[]{
    \includegraphics[width=.3\textwidth]{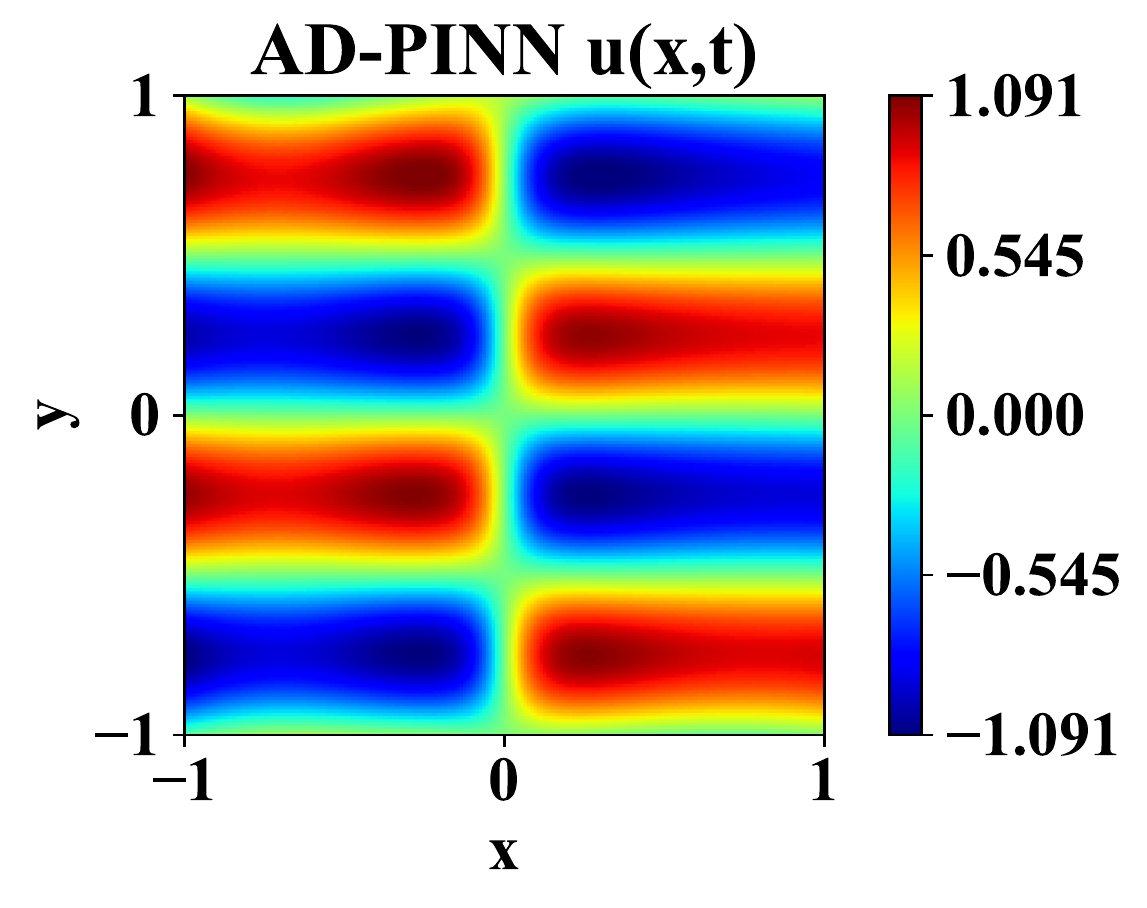}}
    \subfigure[]{
    \includegraphics[width=.3\textwidth]{Poissons2_actual.pdf}}
    \caption{Results of the non-homogeneous Poisson’s example with non zero source function with 4225 collocation points; (a) Solution predicted by the SP-PINN, (b) AD-PINN, (c) Actual solution}
    \label{fig:case43}
\end{figure}

\begin{figure}[!h]
    \centering
    \subfigure[]{
    \includegraphics[width=.3\textwidth]{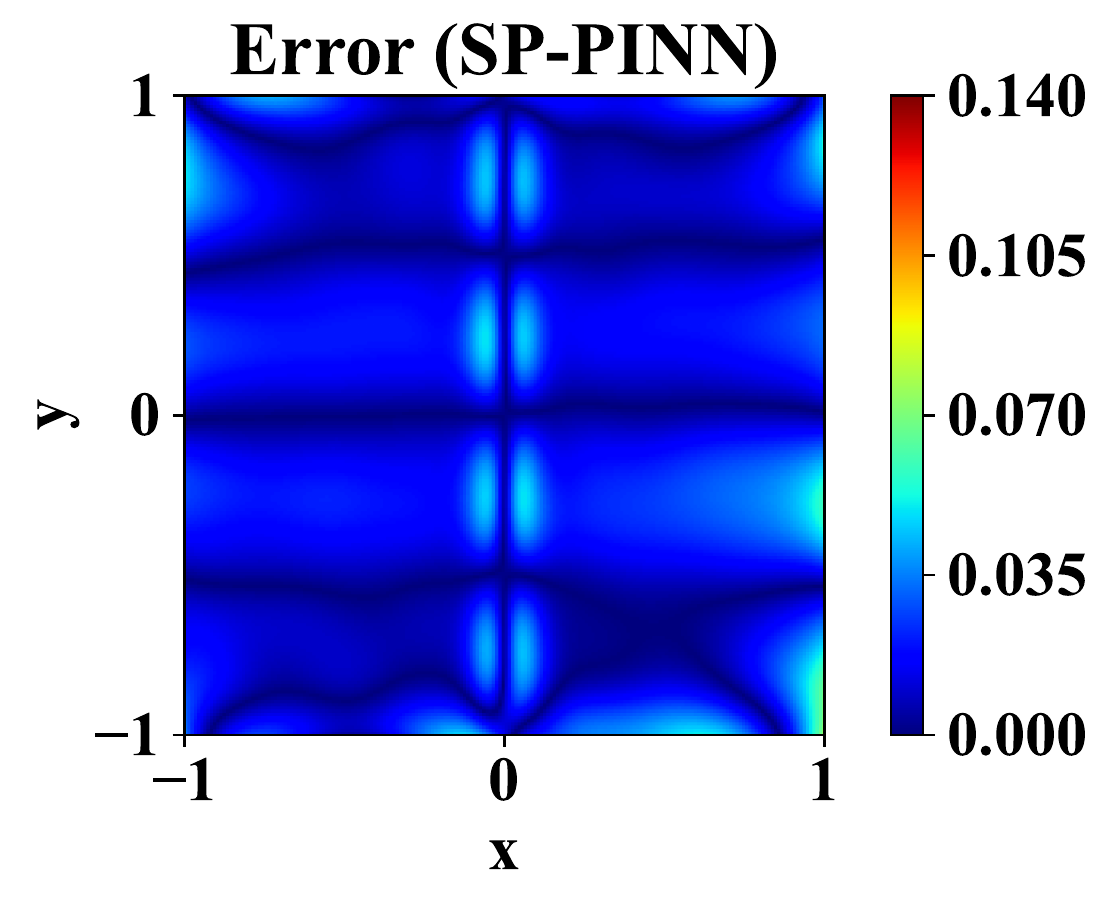}}
    \subfigure[]{
    \includegraphics[width=.3\textwidth]{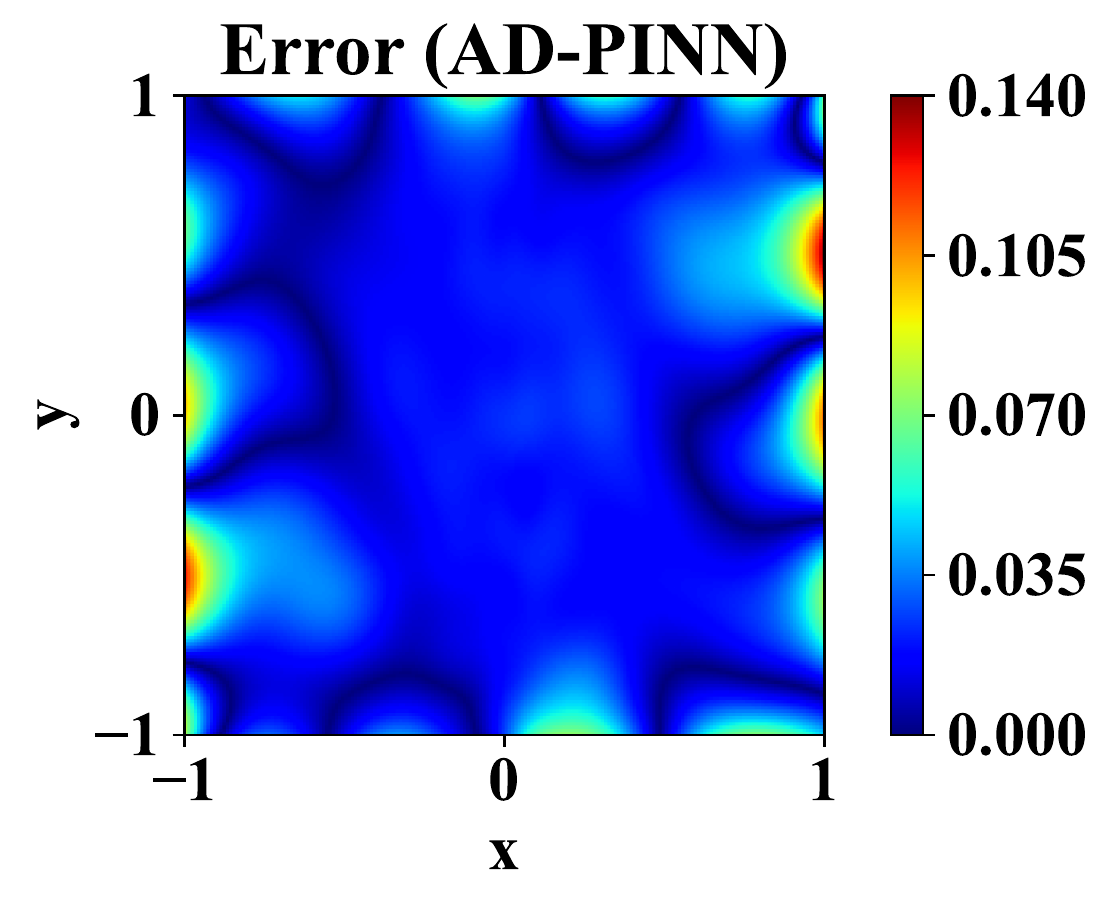}}
    \caption{Prediction error of non-homogeneous Poisson’s example with non zero source function with 4225 collocation points; (a) The absolute error of SP-PINN. (b) The absolute error of AD-PINN}
    \label{fig:case44}
\end{figure}

\begin{table}[ht!]
    \centering
    \caption{Variation of prediction error with number of  collocation points for for the case of 2-D Poisson’s example with non zero source function}
    \label{Table_Poissons22}
\begin{tabular}{lccccc} 
\hline
\multirow{2}{*}{\textbf{Number of collocation points}} &  \multicolumn{2}{c}{\textbf{Max. absolute  error}} &  \multicolumn{2}{c}{\textbf{Average error}}\\
 & {SP-PINN} & {AD-PINN} & {SP-PINN} & {AD-PINN} \\\hline
2601 &	0.0727 & 0.1368 & 0.0145 & 0.0219\\
4225 &	0.0457 & 0.213  & 0.0104 & 0.0199\\
\hline  
\end{tabular}
\end{table}
\subsubsection{Poisson equation over L-shaped domain}
\begin{figure}[!h]
    \centering
    \subfigure[]{
    \includegraphics[width=.32\textwidth]{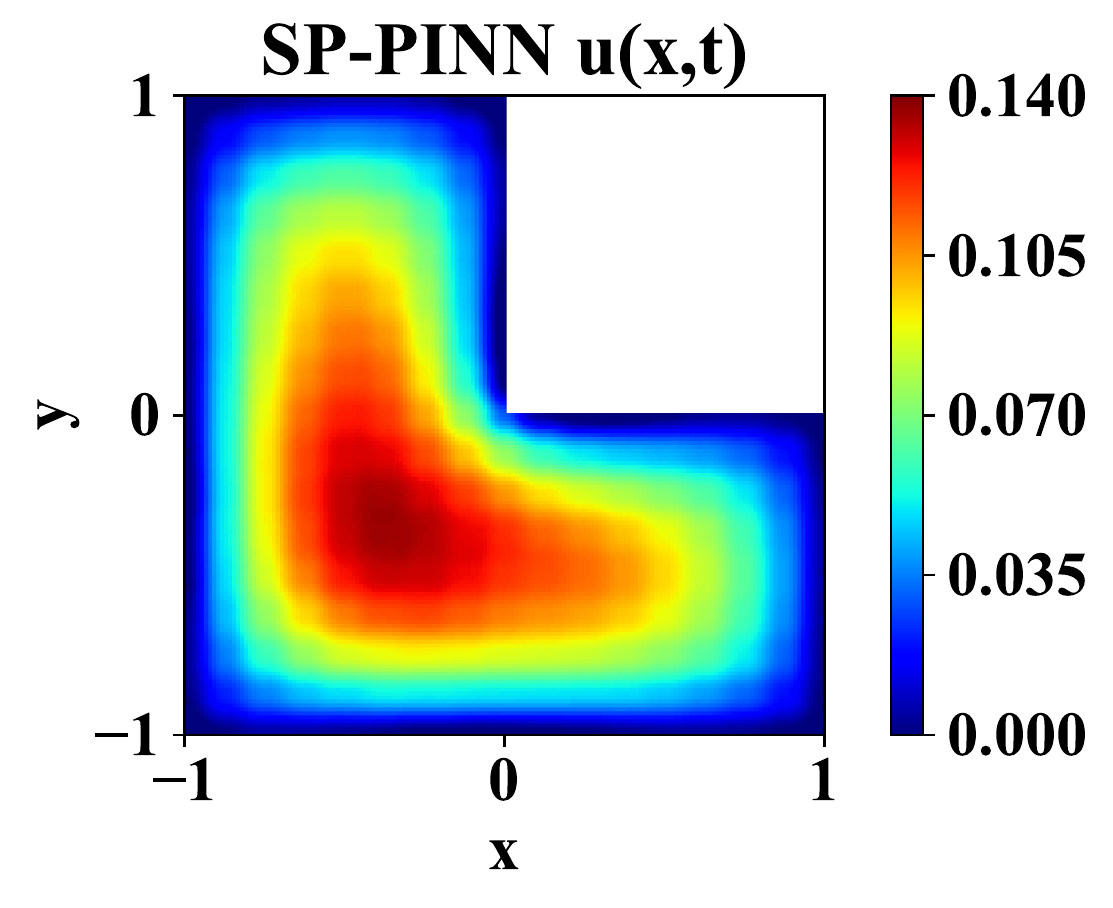}}
    \subfigure[]{
    \includegraphics[width=.32\textwidth]{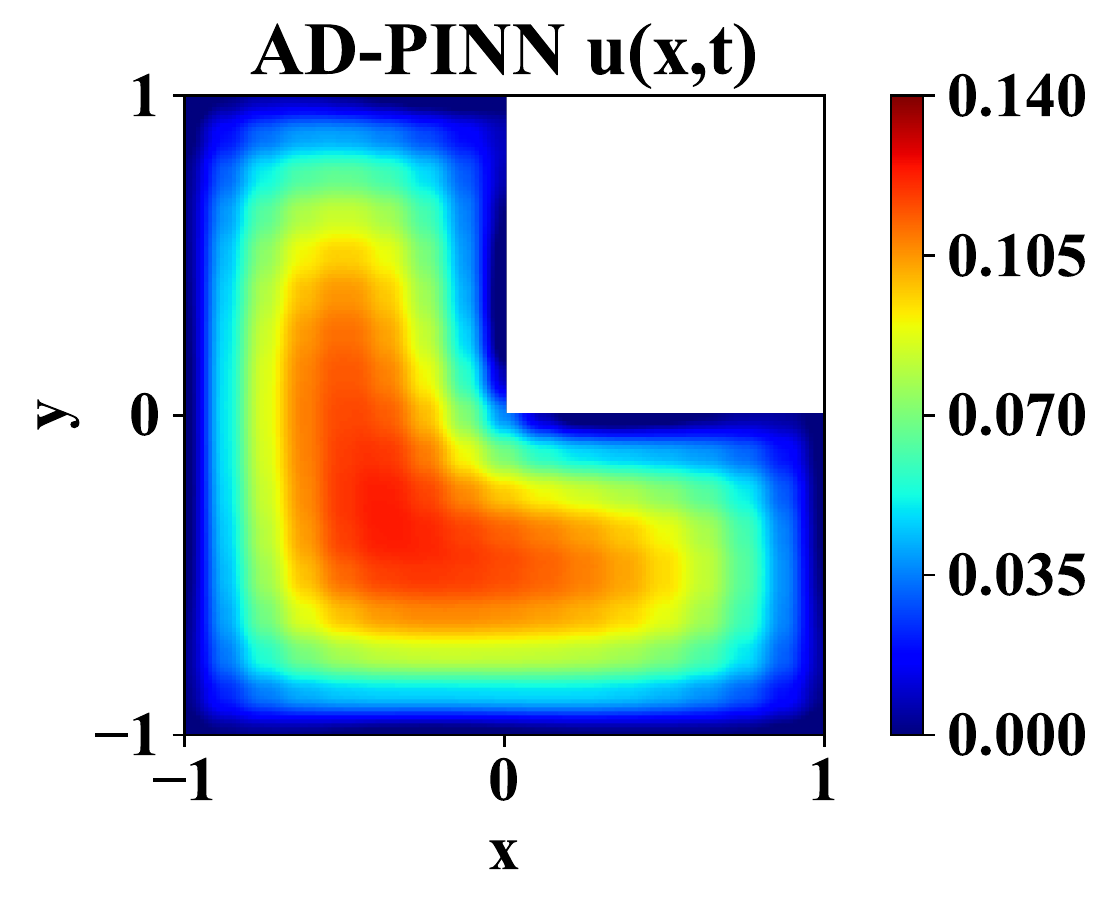}}
    \subfigure[]{
    \includegraphics[width=.32\textwidth]{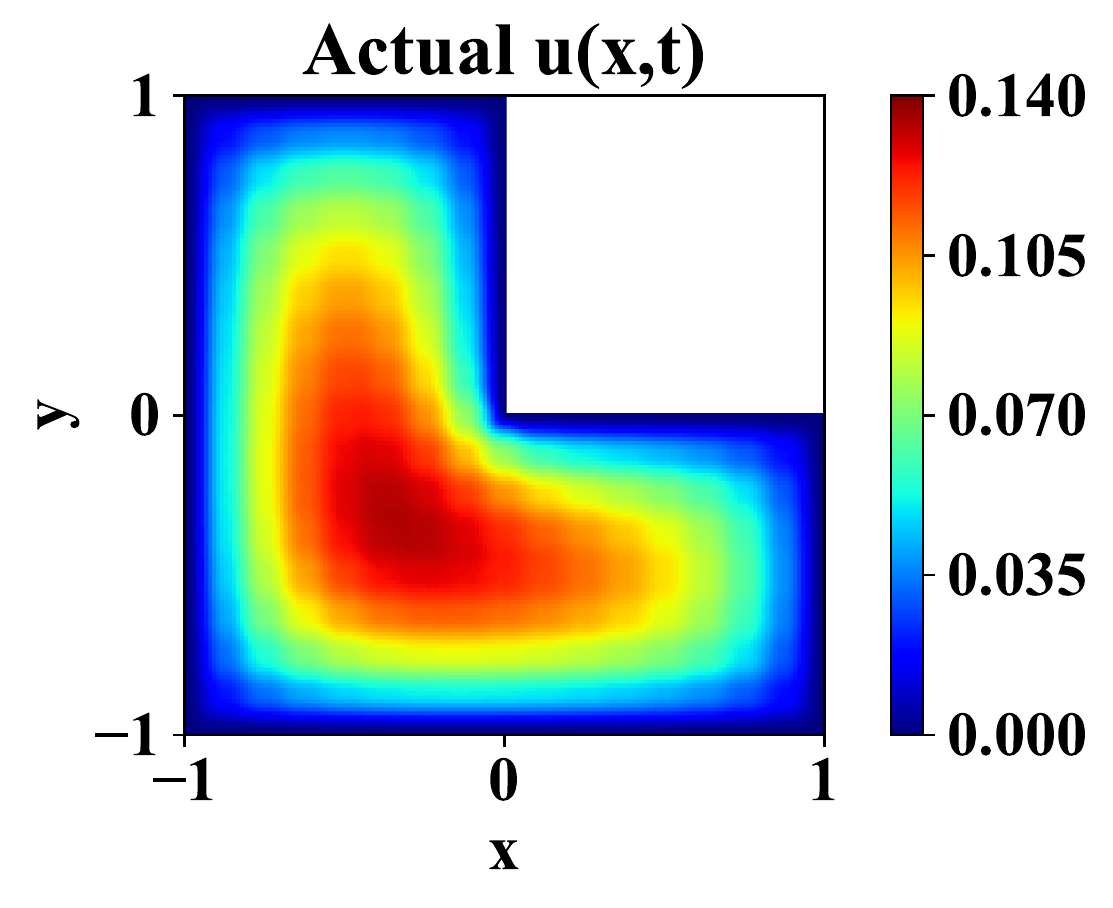}}
    \caption{Results of Poisson equation on L-shaped domain; a) Solution
    predicted by the SP-PINN, (b) AD-PINN, (c) Actual solution}
    \label{fig:case51}
\end{figure}

\begin{figure}[!h]
    \centering
    \subfigure[]{
    \includegraphics[width=.32\textwidth]{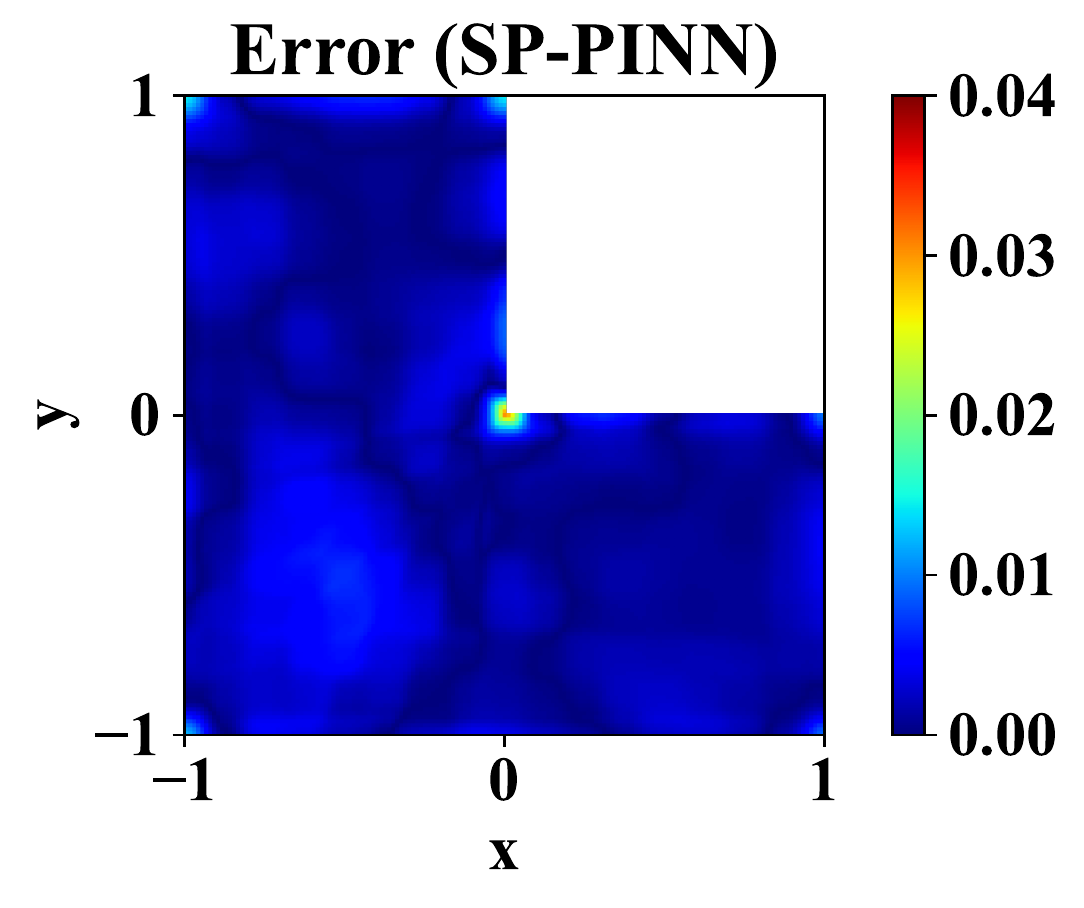}}
    \subfigure[]{
    \includegraphics[width=.32\textwidth]{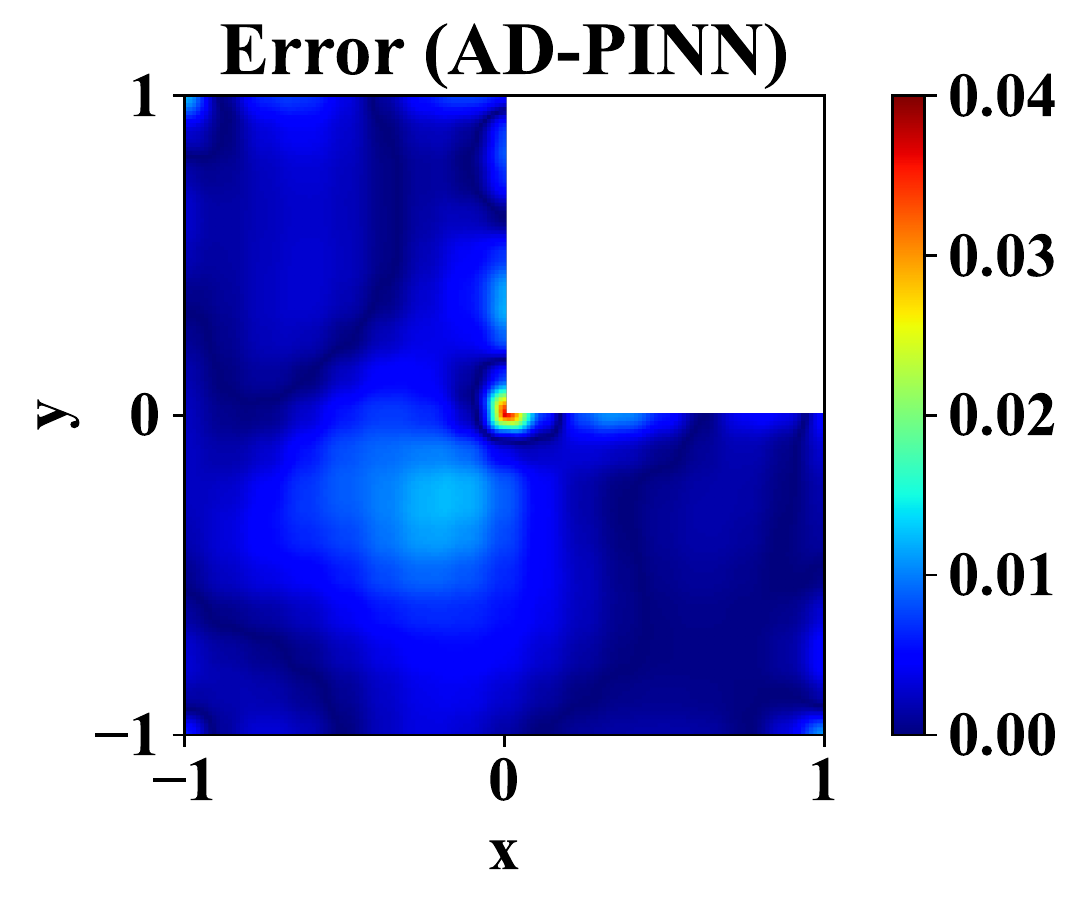}}
    \caption{Prediction error of Poisson equation on L-shaped domain; (a) The
    absolute error of SP-PINN. (b) The absolute error of AD-PINN}
    \label{fig:case52}
\end{figure}

Next, we consider Poisson's equation over an L-shaped domain \cite{karniadakis2005spectral}. The governing equation again remains same as in \autoref{Poissons equation1}. For this example, we have considered
\[f(\bm x) = 1 \]
The objective here is to solve the Poisson's equation over the L-shaped domain. The network architecture used for this problem is identical to \autoref{comp1} with the only difference residing in the fact that ELU activation function has been used here. Optimizer setup is same as \autoref{comp1}, and 1935 collocation points have been used
\autoref{fig:case51} depicts the results obtained using SP-PINN, AD-PINN, and the actual solution (ground truth). We observe that SP-PINN yields superior result as compared to AD-PINN. More specifically, AD-PINN underestimates the peak response and fails to capture the response near the $x=y=0$. This is also evident from the error plot shown in \autoref{fig:case52}. For quantitative assessment, maximum absolute error and average error are shown in table \autoref{tabular-shape_5}. 
\begin{table}[ht!]
    \centering
    \caption{Comparison of prediction error for Poisson's example on L-shaped domain}
    \label{tabular-shape_5}
\begin{tabular}{lcccc} 
\hline
\textbf{Method}&\textbf{Number of collocation points} & \textbf{Max. absolute  error} & \textbf{Average error}\\\hline
SP-PINN & 1935 &	0.03  & 0.002 \\
AD-PINN & 1935 & 0.04  & 0.003 \\
\hline 
\end{tabular}
\end{table}
\subsubsection{Poisson's equation over star-shaped domain}
We next considered the Poisson's example on the star-shaped domain \cite{xiang2022hybrid}. This problem has an analytical solution that takes the following form:
\begin{equation}
    u(x,y) = exp(-({2}x^2+{4}y^2))+\frac{1}{2}
\end{equation}
The objective here is to investigate performance of SP-PINN in solving this problem. The challenge in this case also resides in the irregular problem domain.

\begin{figure}[!h]
    \centering
    \subfigure[]{
    \includegraphics[width=.32\textwidth]{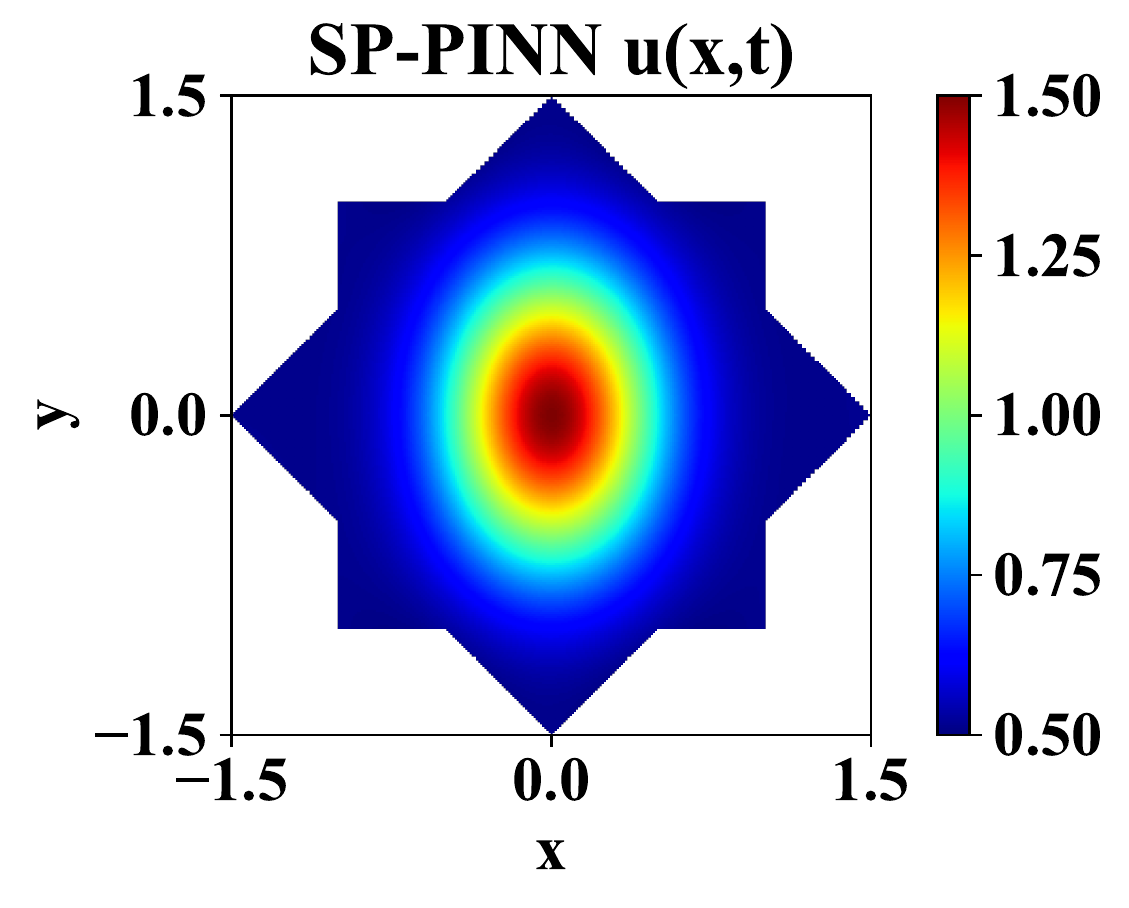}}
    \subfigure[]{
    \includegraphics[width=.32\textwidth]{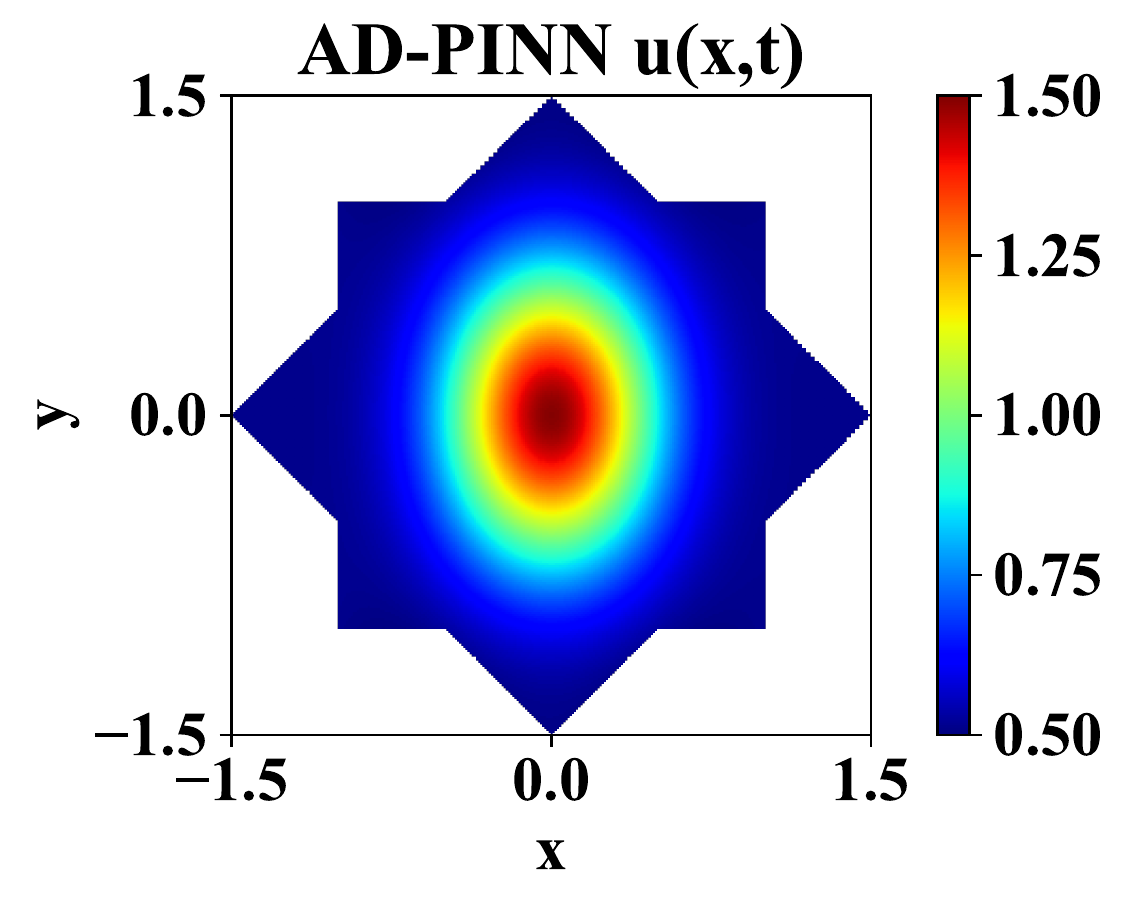}}
    \subfigure[]{
    \includegraphics[width=.32\textwidth]{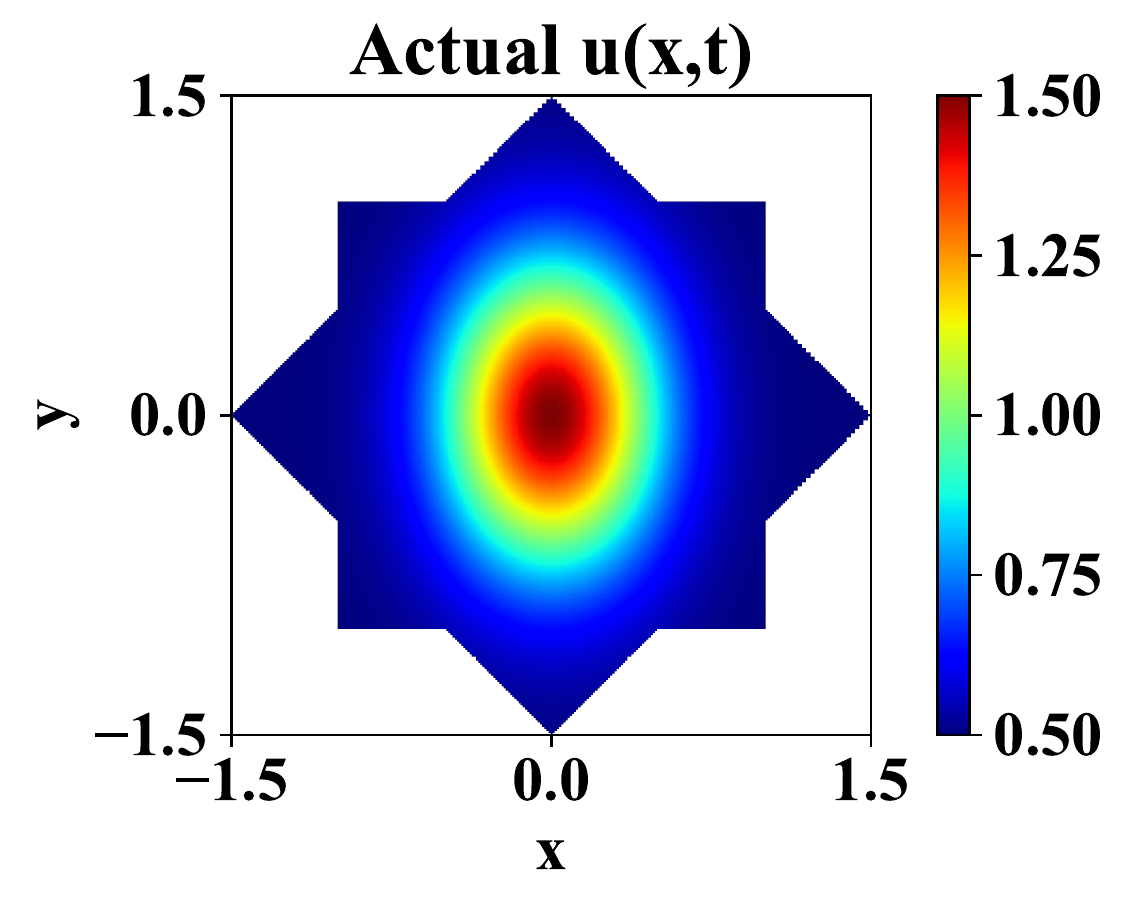}}
    \caption{Poisson equation on L-shaped domain; a) Solution
    predicted by the SP-PINN, (b) AD-PINN, (c) Actual solution}
    \label{fig:case61}
\end{figure}

\begin{figure}[!h]
    \centering
    \subfigure[]{
    \includegraphics[width=.32\textwidth]{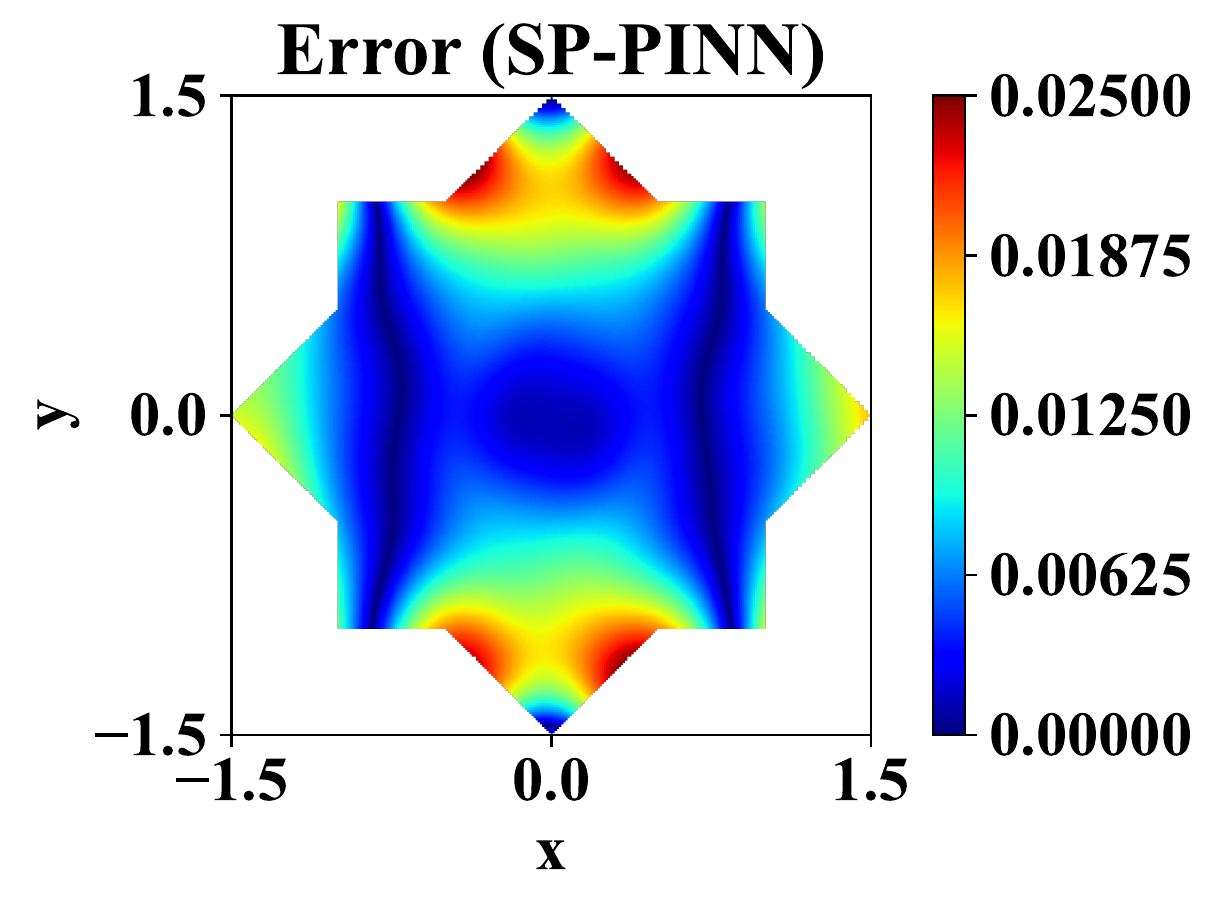}}
    \subfigure[]{
    \includegraphics[width=.32\textwidth]{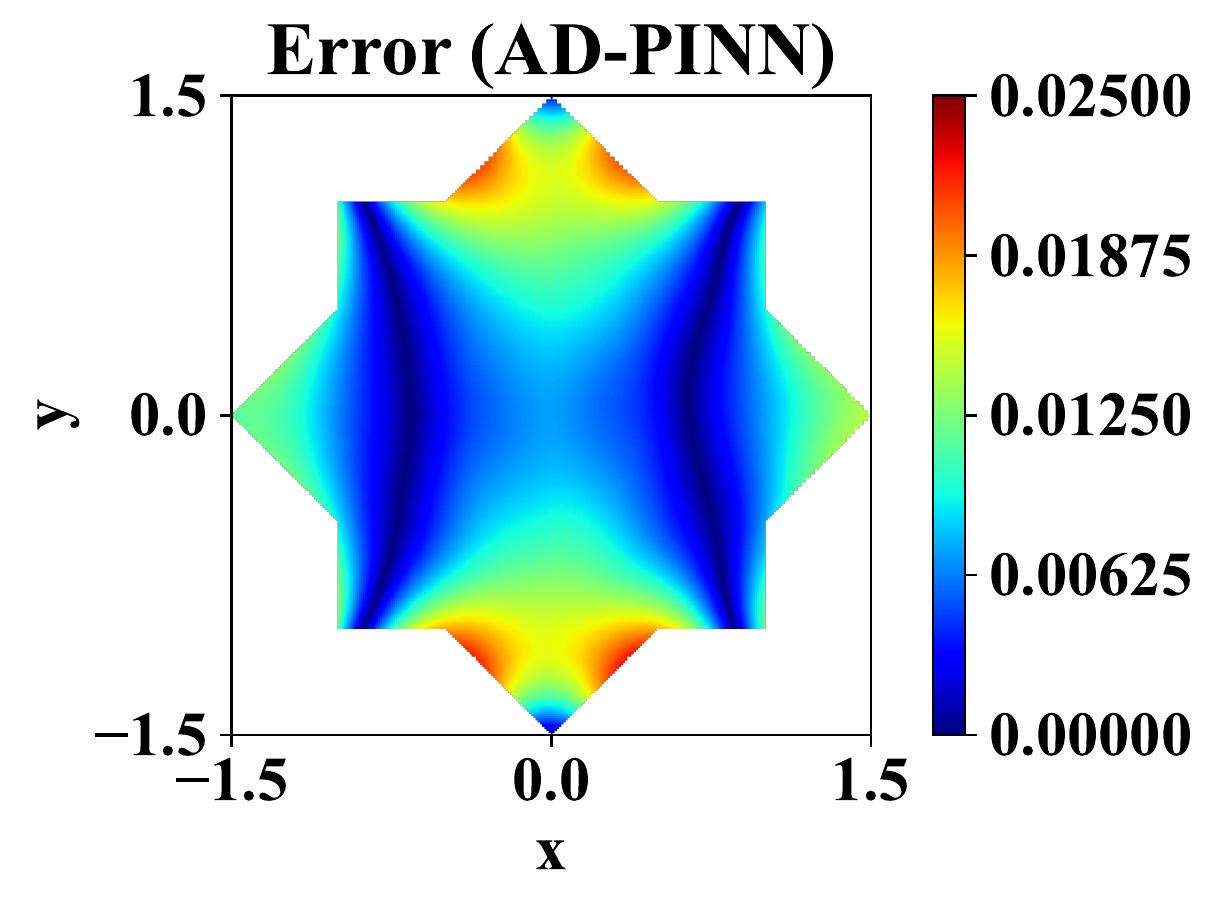}}
    \caption{Poisson equation on L-shaped domain; (a) The
    absolute error of SP-PINN. (b) The absolute error of AD-PINN}
    \label{fig:case62}
\end{figure}

We generate 4595 collocation points inside the geometry for training the framework. A deep fully connected network architecture with the learning rate and the training setup identical to \autoref{comp1} is employed here to obtain the solution. Results obtained using SP-PINN, AD-PINN, and the reference solutions are presented in \autoref{fig:case61}, while the error plots are showcased in \autoref{fig:case62}.
The maximum absolute errors and the average absolute errors for SP-PINN and AD-PINN are listed in the \autoref{Tabular_doublesquare_6}. 
Overall, the proposed SP-PINN yields superior result as indicated by the average error. However, for this problem, SP-PINN is less accurate in capturing the peak response and this is reflected in the maximum absolute error reported in Table \autoref{Tabular_doublesquare_6}
The noteworthy observation here is that although the maximum absolute error of the SP-PINN is slightly more than that of the AD-PINN, the average error of the SP-PINN is less than that of the AD-PINN.
\begin{table}[h]
    \centering
    \caption{Comparison of prediction error for Poisson's example on star -shaped domain}
    \label{Tabular_doublesquare_6}
\begin{tabular}{lcccc} 
\hline
\textbf{Method}&\textbf{Number of collocation points} & \textbf{Max. absolute  error} & \textbf{Average error}\\\hline
SP-PINN & 4595 &  0.02483  & 0.00763 \\
AD-PINN & 4595 &  0.02233  & 0.00793 \\
\hline 
\end{tabular}
\end{table}
\subsection{Fourth-order phase field fracture problem}
\begin{figure}[!h]
    \centering
    \includegraphics[width=.5\textwidth]{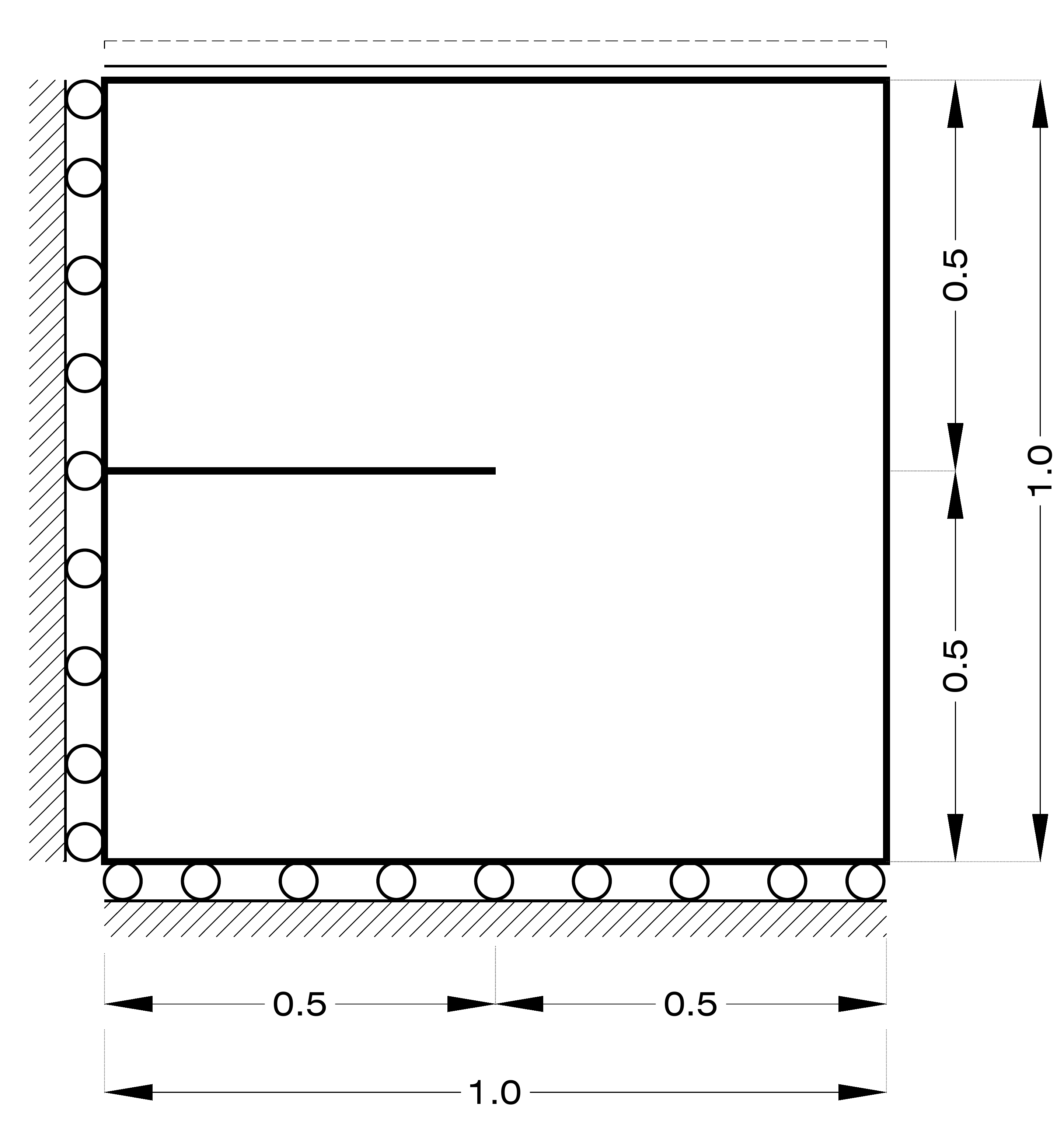}
    \caption{The geometry and boundary conditions for single notched plate specimen. The units of the dimensions are in mm}
    \label{specimen1}
\end{figure}

As the last example, we consider the fracture of a square plate with a horizontal crack spanning from the midpoint to the outer left edge. For solving a fracture problem, the phase field modeling \cite{miehe2010phase} approach is utilized. The geometry and boundary conditions  for the problem are depicted in the \autoref{specimen1}. Moreover, the material parameters are considered to be $\lambda=121.15 \text{kN/mm}$, $\mu = 80.77 \text{kN/mm}^{2}$ and $G_c = 2.7\times10^{-3}\text{KN/mm}$. A step-wise increment in displacement, $\delta{u} = 1\times10^{-3}\text{mm}$ is applied to simulate the crack propagation.
\par
An energy-based approach is employed to obtain the numerical solution for the  problem. Since the method is based on the concept of energy conservation, it is necessary to compute the  total energy of the system. The system under consideration is a linear elastic system here. Thus, the total elastic strain energy of a linear elastic body is computed as the sum of the elastic strain energy stored in the system $\Psi_{e}(\epsilon)$ and the work done by external forces. the total energy, $mathcal{E}$ can be expressed as:
\begin{equation}
  \mathcal{E}=\Psi_e(\epsilon)-W_{ext}.
\end{equation}
While $\Psi_e(\in)$ is expressed in terms of components of linearized strain tensor, $\in(u)$, the external work done, $W_{\text {ext }}$ formulated in terms of surface traction, $t_N$ and the body forces, $f$ and displacement $u$. From the total energy of the linear elastic system, equilibrium equations can be achieved based on the variational energy. The displacement field, $u$ that minimizes the energy function $\mathcal{E}$ obtains the equilibrium equation in terms of Cauchy stress ($\sigma$) as:
\begin{equation}\label{eq:equilibrium}
-\nabla \cdot \sigma = f \hspace{2em} \text{on}  \ \Omega
\end{equation}
\autoref{eq:equilibrium} satisfies the following Neumann and Dirichlet boundary conditions:
\begin{equation}
\begin{aligned}
\sigma \cdot \boldsymbol{n} &=\boldsymbol{t}_N \text { on } \partial \Omega_N \\
\boldsymbol{u} &=\overline{\boldsymbol{u}} \text { on } \partial \Omega_D,
\end{aligned}
\end{equation}
\par
We couple the phase field approach in the energy based formulations. The phase field method is a commonly employed in continuum modelling techniques \cite{miehe2010phase,miehe2010thermodynamically} to solve  both the displacement field and fracture region simultaneously by minimizing energy. The approach generalizes Griffith's theory necessitating no assumptions for the growth of cracks \cite{schmidt2009eigenfracture}. One of the noteworthy works which employ Griffith's theory is presented in \cite{bourdin2000numerical}. The work illustrates a variational method to fracture, formulated numerically, in which the sharp crack surface topology in a solid is represented by a diffusive crack zone guided by the length scale parameter $l_0$. In the phase field modeling approach, the propagation of the crack is measured using a continuous scalar-valued phase-field parameter function $\phi(\bm x)\in [0,1]$. The phase field parameter is incorporated into the equilibrium equation of elasticity in terms of a stress-degradation function, $g(\bm x)$, in order to show the weakening of the material in the neighborhood area of the crack. In practice, a commonly used stress-degradation function for isotropic materials is given by  \cite{miehe2010phase}:
\begin{equation}
   g(\phi)=(1-\phi)^2 . 
\end{equation}
With the incorporation of the degradation function, the equilibrium equation can be rewritten as:
\begin{equation}\label{eq:equilibrium1}
-\nabla g(\phi(\bm x)) \cdot \sigma = f, \hspace{2em} \text{on}  \ \Omega.
\end{equation}
Correspondingly, the elastic field is constrained by Dirichlet and Neumann boundary conditions:
\begin{equation}
  \begin{aligned}
   g(\phi) \sigma \cdot \boldsymbol{n} &=\boldsymbol{t}_N \text { on } \partial \Omega_N \\
   \boldsymbol{u} &=\overline{\boldsymbol{u}} \text { on } \partial \Omega_D,
\end{aligned}  
\end{equation}
where $\bold{t}_N$ is the prescribed boundary forces and $\overline{\boldsymbol{u}}$ is the prescribed displacement for each load step. The Dirichlet and Neumann boundaries are represented by $\partial \Omega_D$ and $\partial \Omega_N$, respectively.
The governing equation for the phase-field, on the other hand, is denoted by:
\begin{equation}\label{eq:pf}
    \frac{G_c}{2}\left[\frac{\phi}{l_0}+\frac{l_0}{2}|\nabla \phi|^2+\frac{l_0^3}{16}(\Delta \phi)^2\right]= -g(\phi) H(\bm x, t) \text { on } \Omega
\end{equation}
where $H(\bm{x}, t)$ is the strain-history function and $G_{c}$ denotes the critical energy release rate which depends on the material. Here, it should be noted that only the tensile component of the major stress degrades as a result of the crack growth, whereas the compressive component does not degrade \cite{miehe2010phase}. Therefore, the total strain energy functional $(\Psi_{0})$ is divided into the strain energy functional due to tensile component $(\Psi_{0}{+})$  and that of compressive component $\left(\Psi_{0}{-}\right)$  as follows: 
\begin{equation}
    \Psi_{0}(\epsilon)= \Psi_{0}^{+}(\epsilon){+}\Psi_{0}^{-}(\epsilon)
\end{equation}
\autoref{eq:pf}, governing propagation of the phase field, has a local strain-history functional, $H(\bm x,t)$, which couples the elastic equation with the phase field equation. $H(\bm x,t)$ is evaluated as the maximum positive tensile energy, $\Psi_0^{+}$ and is mathematically expressed as
\begin{equation}
    H(\bm x, t)=\max _{s \in[0, t]} \Psi_0^{+}(\in(\bm x, s)),
\end{equation}
where $\bm x$ is the integration point. The local strain-history functional approach is used  here to define initial cracks in the system \cite{miehe2010phase}; this  guarantees monotonically increasing values $\phi$ and thus hinder the crack from healing \cite{miehe2010thermodynamically}. The initial strain-history function $H(\bm x, 0)$ can be expressed as a function of the closest distance between any point $\bm x$ on the domain and the line $l$ representing discrete crack \cite{borden2012phase}. The expression of the initial history function can be represented by:
\begin{equation}\label{init_hist}
    H(\bm x, 0)=\left\{\begin{array}{ll}
\frac{B G_c}{2 l_0}\left(1-\frac{2 d(\bm x, l)}{l_0}\right) & d(\bm x, l) \leqslant \frac{l_0}{2}, \\
0 & d(\bm x, l)>\frac{l_0}{2}
\end{array},\right.
\end{equation}
In \autoref{init_hist}, $B$ is a scalar parameter that influences the magnitude of the scalar history field, and is given  as:
\begin{equation}
    B=\frac{1}{1-\phi} \text { for } \phi<1.
\end{equation}
For the fourth-order phase field model, the crack density functional is defined as:
\begin{equation}\label{crack_density}
    \Gamma_4(\phi)=\frac{1}{2 l_0} \int_{\Omega}\left(\phi^2+\frac{l_0^2}{2}|\nabla \phi|^2+\frac{l_0^4}{16}(\Delta \phi)^2\right) d \Omega .
\end{equation}
The surface energy of a newly developed crack is given by \cite{molnar20172d}: 
\begin{equation}\label{surf_energy_n}
    \Psi_c=\int_{\Omega}\left[G_c \Gamma_n(\phi)+g(\phi) H(x, t)\right] d \Omega.
\end{equation}
The incorporation of the \autoref{crack_density} in \autoref{surf_energy_n} results in the expression of the  surface energy,
\begin{equation}
    \Psi_c  =\int_{\Omega}\left(\frac{G_c}{2}\left[\frac{\phi}{l_0}+\frac{l_0}{2}|\nabla \phi|^2+\frac{l_0^3}{16}(\Delta \phi)^2\right]+g(\phi) H(x, t)\right) d \Omega .
\end{equation}
The presumption made here for the phase field is that it satisfies homogeneous Neumann-type boundary conditions on the boundary:
\begin{equation}
    \begin{array}{ll}
\Delta \phi & =0 \text { on } \partial \Omega, \\
\nabla\left(l_0^4 \Delta \phi-2 l_0^2 \phi\right) \cdot n & =0 \text { on } \partial \Omega .
\end{array}
\end{equation}
The total elastic strain energy due to the fracture is formulated as:
\begin{equation}
    \Psi_e  =\int_{\Omega}\left(g(\phi) \Psi_0^{+}(\boldsymbol{\epsilon})+\Psi_0^{-}(\boldsymbol{\epsilon})\right) d \Omega
\end{equation}
The problem statement for the phase field fracture employing a variational energy approach can be stated as:
\begin{equation}
\begin{array}{ll}
\text { Minimize: } \quad \mathcal{E} & =\Psi_e+\Psi_c, \\
\text { subject to: } \quad u & =\bar{u} \text { on } \partial \Omega_D
\end{array}
\end{equation}
It is noteworthy that in the variational energy form, the homogeneous Neumann boundary conditions are implicitly satisfied. 
\par
\begin{figure}[!h]
    \centering
    \subfigure[]{
    \includegraphics[width=.32\textwidth]{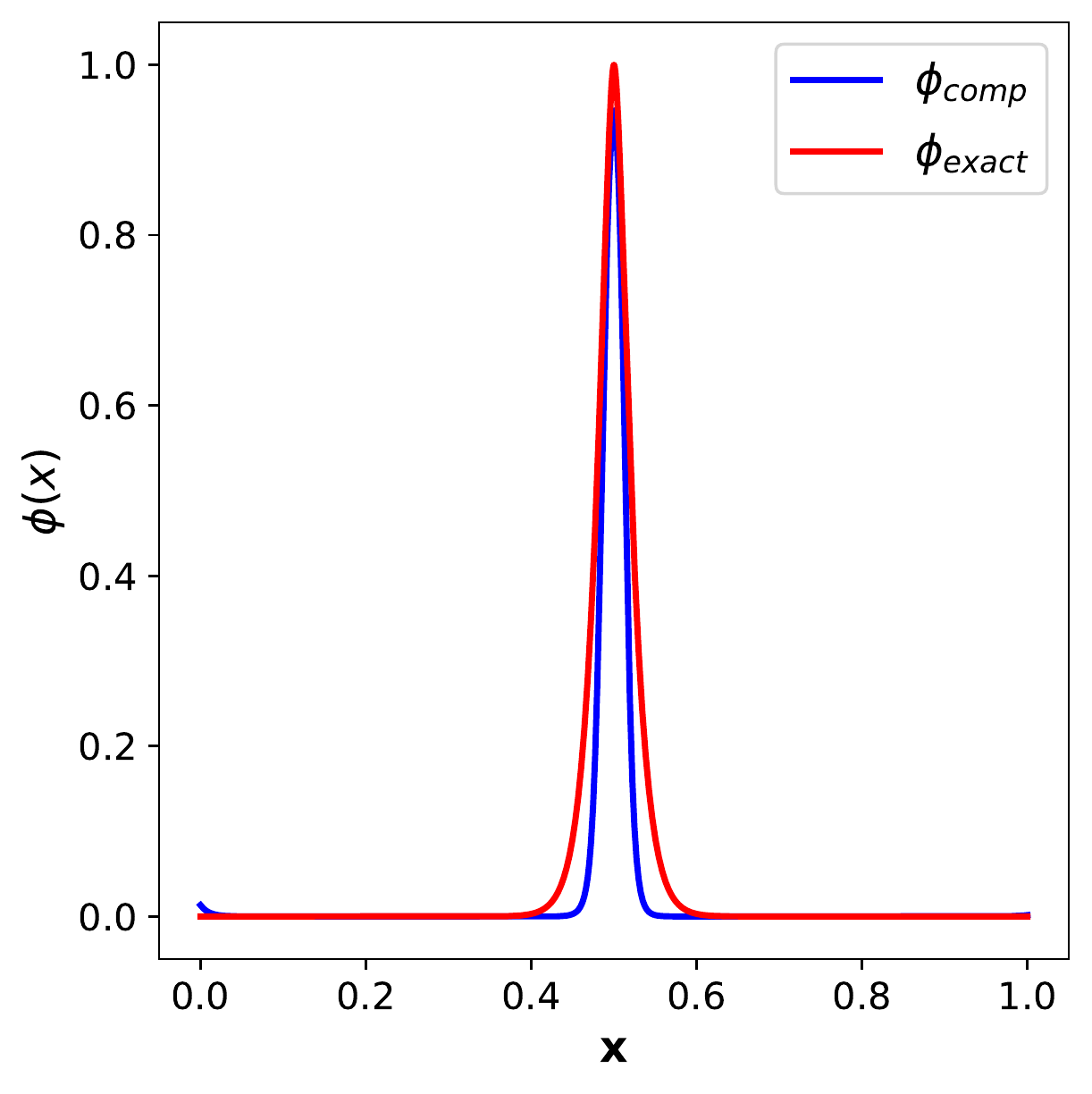}}
    \subfigure[]{
    \includegraphics[width=.32\textwidth]{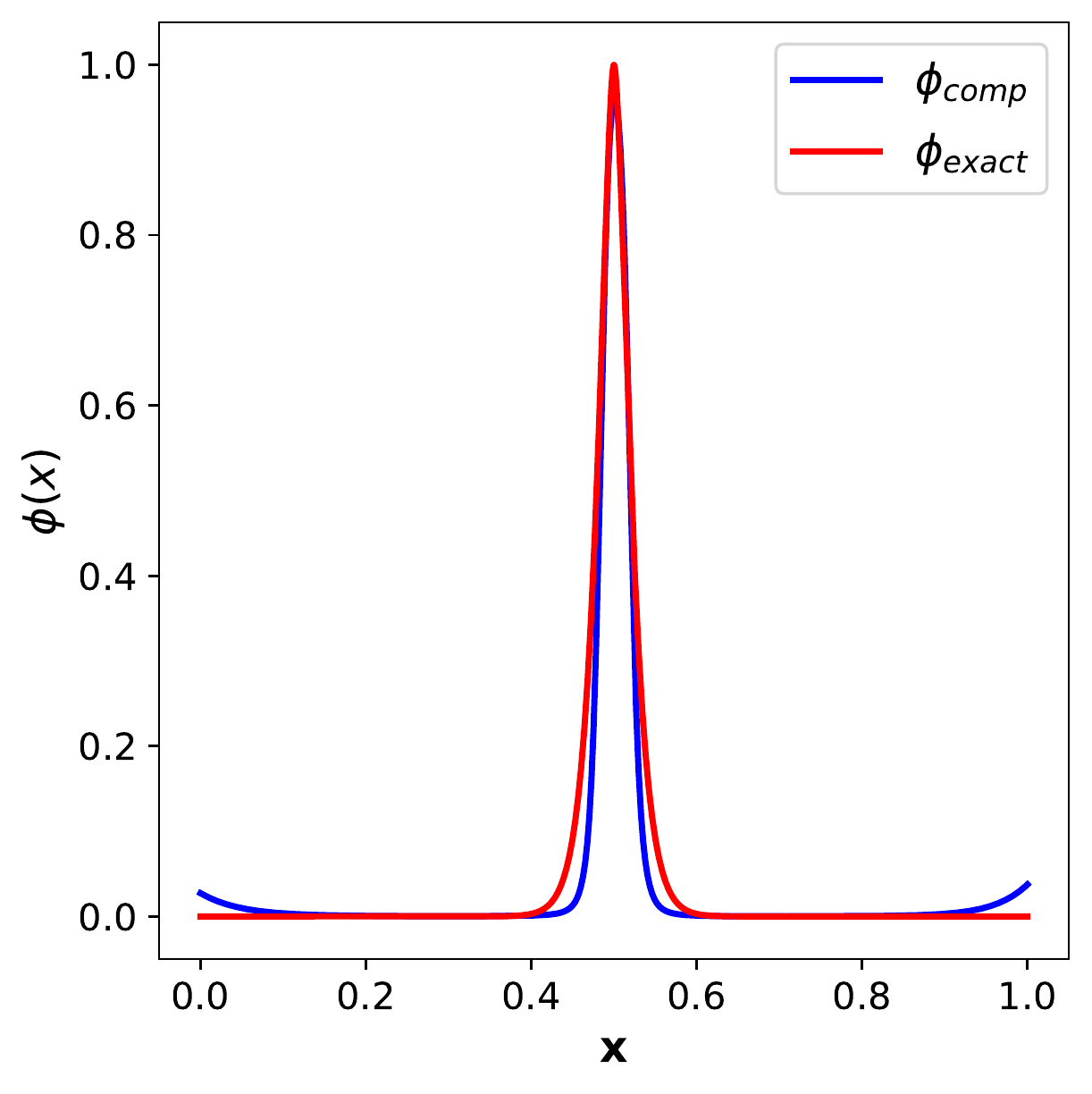}}
    \subfigure[]{
    \includegraphics[width=.32\textwidth]{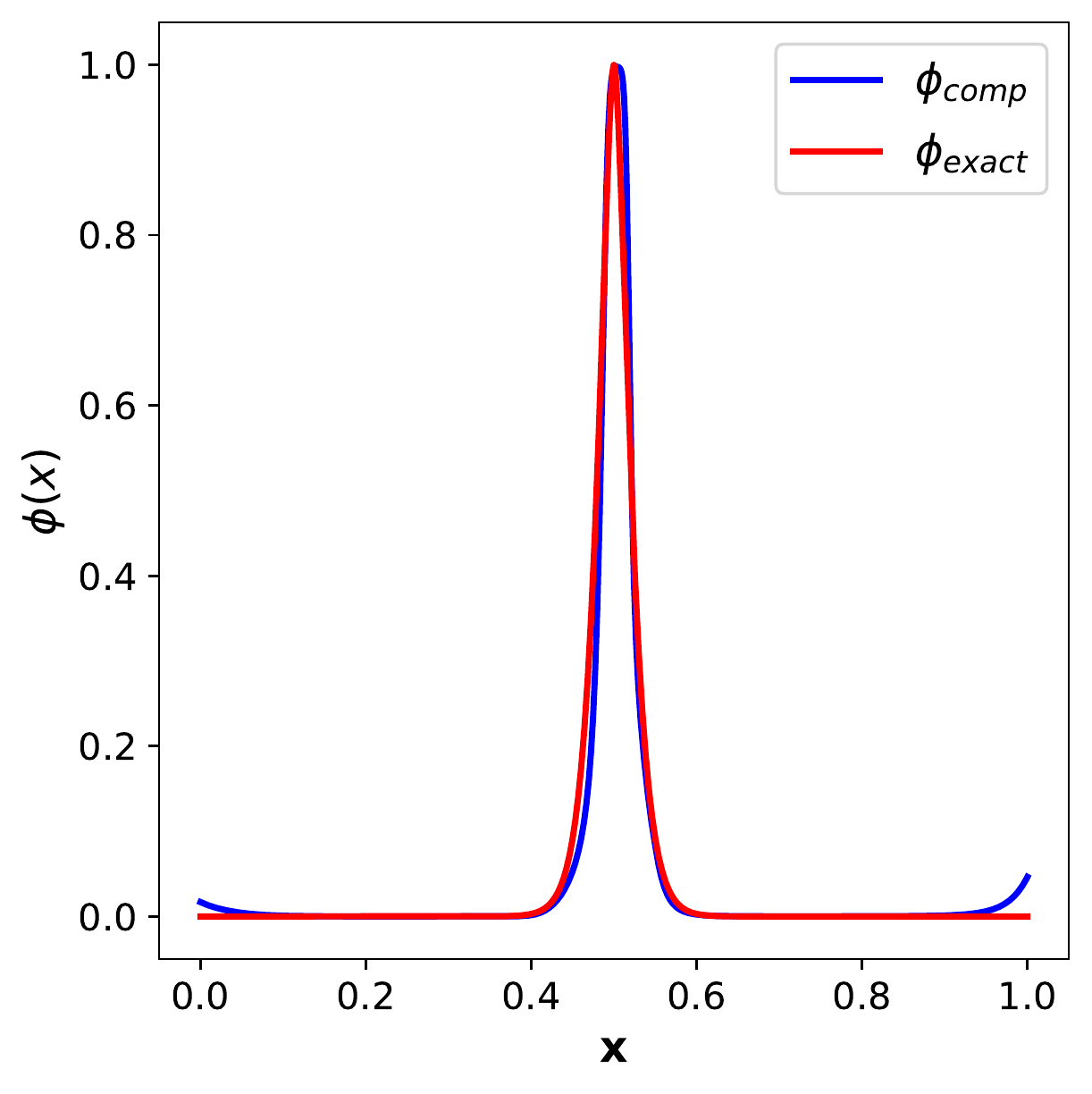}}
    \caption{Variation of Phase field $\Phi$ a) crack length $2\times10^{-3}$, (b) crack length $4\times10^{-3}$ (c) $6\times10^{-3}$}
    \label{fig:Phasefield}
\end{figure}
\begin{figure}[!h]
    \centering
    \subfigure[]{
    \includegraphics[width=.32\textwidth]{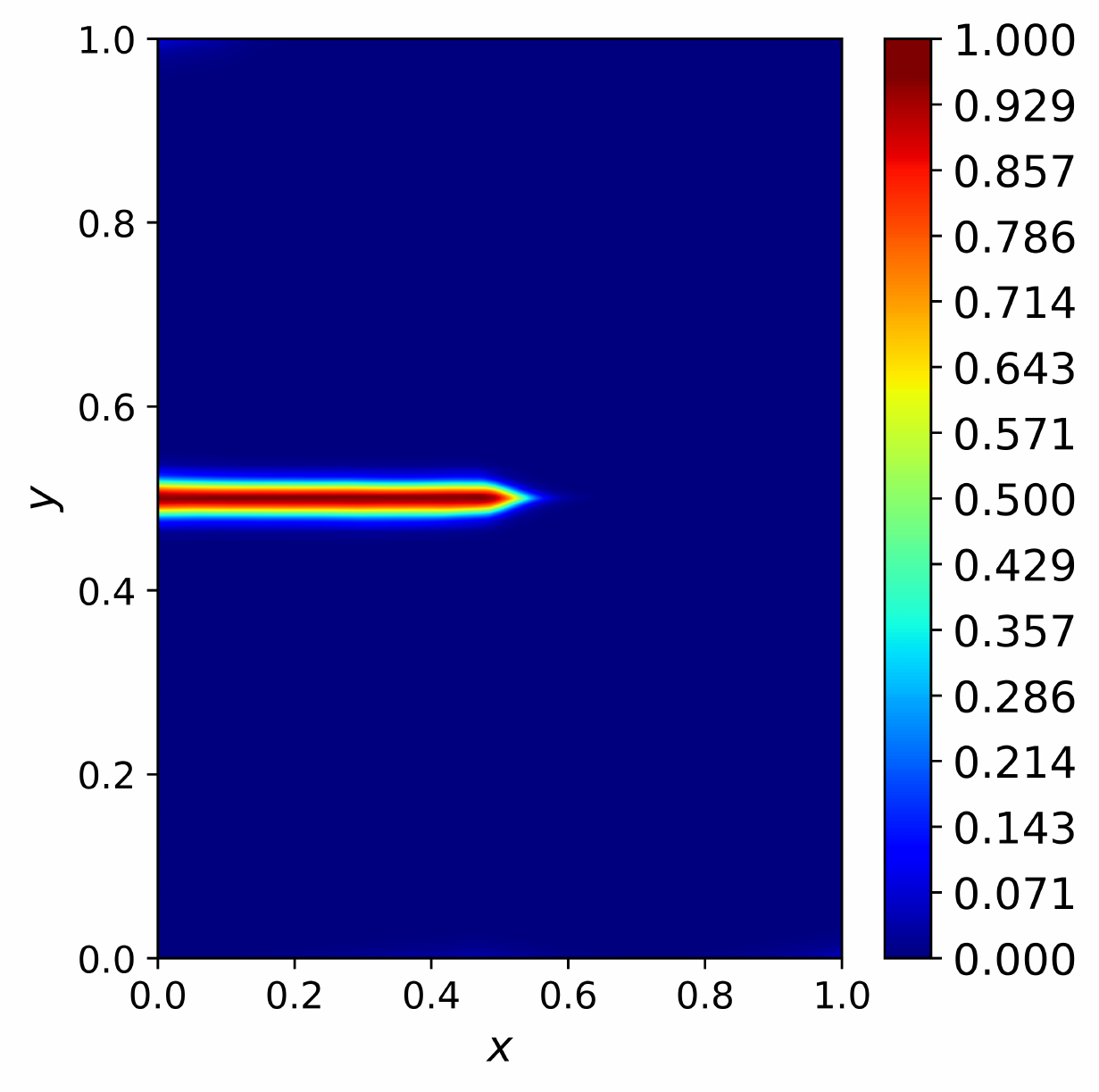}}
    \subfigure[]{
    \includegraphics[width=.32\textwidth]{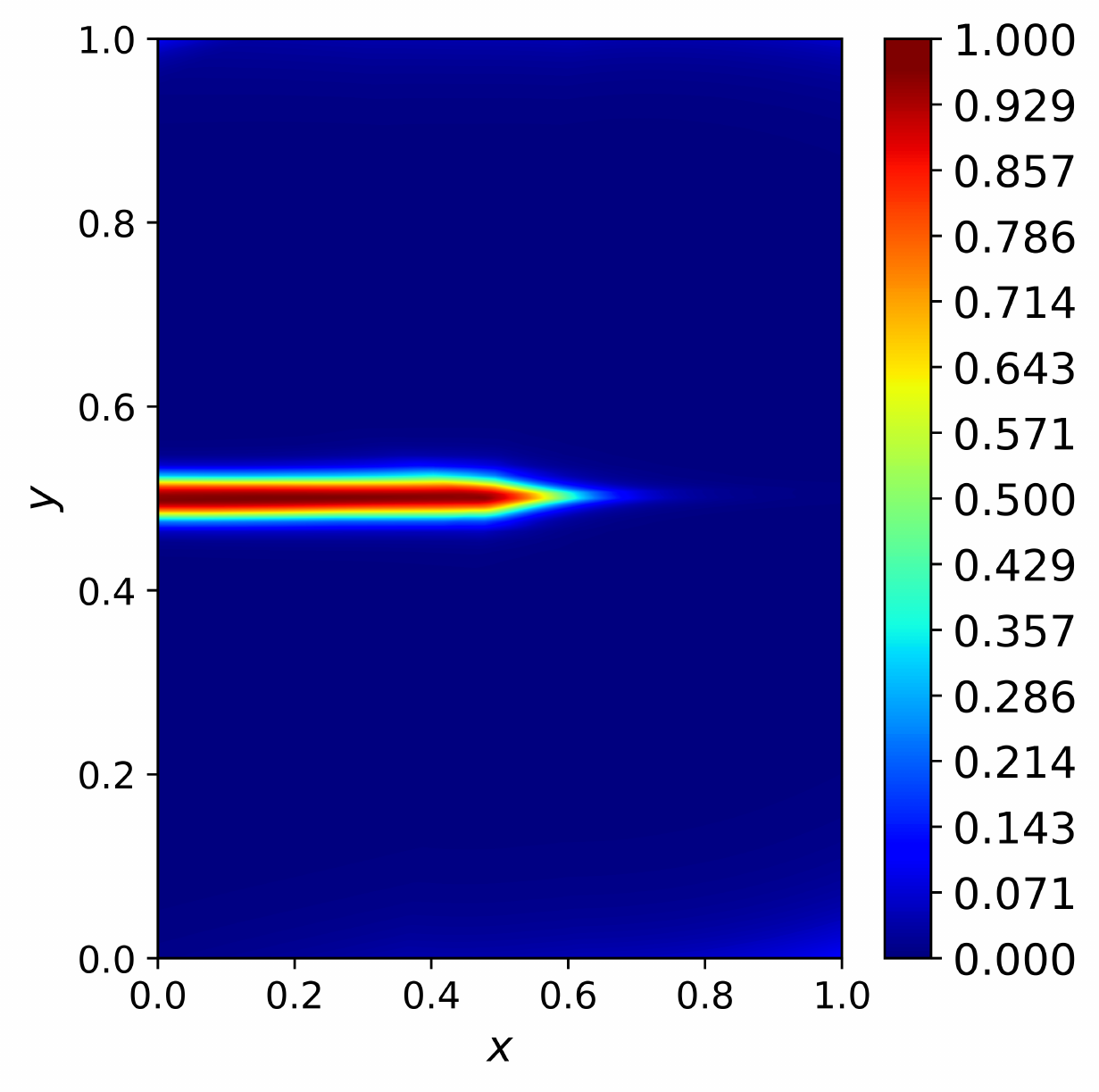}}
    \subfigure[]{
    \includegraphics[width=.32\textwidth]{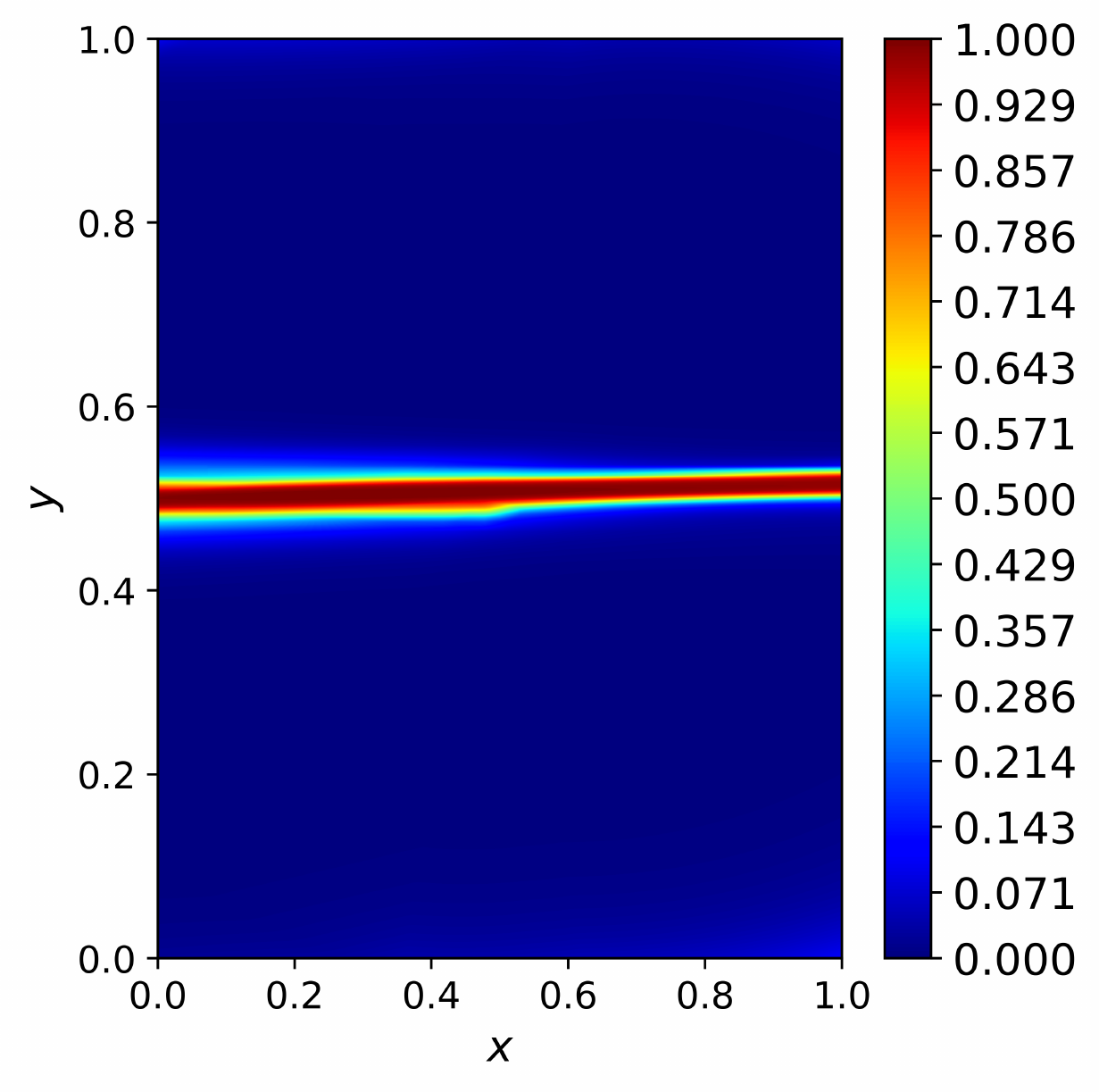}}
    \caption{Evolution of the crack; a) Solution
    predicted by the SP-PINN, (b) AD-PINN, (c) Actual solution}
    \label{fig:increment of crack length}
\end{figure}

Now, to obtain SP-PINN-based solution, we use a fully connected network architecture with 4 hidden layers having 50 neurons in each. \texttt{tanh} activation function is used in the first, second, and fourth hidden layers. For the third hidden layer, we have used the SELU activation function. The output layer has a linear activation function. In addition, the length-scale parameter in \autoref{init_hist} is chosen to be, $l_0= 0.125$. The displacement boundary conditions are given by:
\begin{equation}
    u(0, y)=v(x, 0)=0, \quad v(x, 1)=\Delta u,
\end{equation}
where $u$ and $v$ are the solutions displacement fields in $x_1$ and $x_2$-axis. On the other hand, the neural network outputs $(\hat{u},\hat{v})$ for the elastic field are modified as follows to incorporate the exact Dirichlet boundary conditions,
\begin{equation}
    \begin{aligned}
&u=[x(1-x)] \hat{u}, \\
&v=[y(y-1)] \hat{v}+y \Delta u,
\end{aligned}
\end{equation}
We begin with a ${20 \times 20}$ coarse mesh with each element having 16 quadrature points. As to initiate the crack, the strain-history functional, \autoref{init_hist} is used.
Fracture simulations are computationally costly as it requires a very fine mesh to determine the damage zone. Adaptive refinement of the mesh \cite{goswami2020adaptive} is done in each step with the growth of the crack. The progression of the crack in three different predefined displacement steps is shown in the \autoref{fig:increment of crack length}, while the evaluation of the phase field in the corresponding steps is shown in \autoref{fig:Phasefield}. It is observed that the proposed SP-PINN is able to solve the fracture propagation problems quite accurately.

\section{Conclusions}\label{sec:conclusions}
In this paper, we have proposed stochastic projection-based physics-informed neural networks (SP-PINN) and illustrated its application in computational mechanics. The proposed approach blends the stochastic projection theory with the traditional physics-informed neural network, and this eliminates the computational bottleneck of automatic  differentiation associated with vanilla PINN. The key observations and the salient features of the proposed approach are highlighted below:

\begin{itemize}
    \item Stochastic projection enables SP-PINN to be flexible with the choice of architecture. For example, one need not worry about differentiability of activation function before selecting it.
    \item Unlike the conventional PINN, SP-PINN is flexible with overall network architecture. For example, non-differentiable (in space and time) neural network architecture like convolutional neural network can also be used with the proposed framework.
    \item In general, SP-PINN solves the PDEs reasonably well. Nevertheless, the approach is advantageous over the state-of-the-art methods for the cases when the governing physics has non-smooth solutions and the problem has a non-trivial domain.
\end{itemize}

Despite the fact that SP-PINN has proven to be a promising framework for solving PDEs, it has some limitations which need further attention. For example, stochastic projection relies on the distribution of the neighborhood points; this can have an effect on the optimal number of neighboring points and hence, needs further investigation. Generally speaking, network architecture in PINN and in SP-PINN are selected based on trial and error. There is a need to automate this. Future research will be conducted to address these limitations

\section*{Acknowledgements}
NN acknowledges the support received from Ministry of Education in the form of Prime Ministers Research Fellowship. SC acknowledges the financial support of Science and Engineering Research Board (SERB) via grant no. SRG/2021/000467.


\end{document}